\documentclass[12pt,a4paper,titlepage]{article}
\usepackage[applemac]{inputenc}
\usepackage{amssymb}
\usepackage{amsmath}
\usepackage{amsopn}
\newcommand{\LL}{\mathbb{L}}
\newcommand{\CC}{\mathbb{C}}
\newcommand{\DD}{\mathbb{D}}
\newcommand{\BB}{\mathbb{B}}
\newcommand{\MM}{\mathbb{M}}
\newcommand{\NN}{\mathbb{N}}
\newcommand{\PP}{\mathbb{P}}
\newcommand{\RR}{\mathbb{R}}
\newcommand{\ZZ}{\mathbb{Z}}
\newcommand{\frA}{\mathfrak{A}}
\newcommand{\frB}{\mathfrak{B}}
\newcommand{\frp}{\mathfrak{p}}
\newcommand{\frm}{\mathfrak{m}}
\newcommand{\frN}{\mathfrak{N}}
\newcommand{\frR}{\mathfrak{R}}
\newcommand{\frU}{\mathfrak{U}}
\newcommand{\kA}{\mathcal{A}}

\newcommand{\kB}{\mathcal{B}}
\newcommand{\kD}{\mathcal{D}}
\newcommand{\kE}{\mathcal{E}}
\newcommand{\kF}{\mathcal{F}}
\newcommand{\kH}{\mathcal{H}}
\newcommand{\kL}{\mathcal{L}}
\newcommand{\kM}{\mathcal{M}}
\newcommand{\kO}{\mathcal{O}}
\newcommand{\kP}{\mathcal{P}}
\newcommand{\kQ}{\mathcal{Q}}
\newcommand{\kS}{\mathcal{S}}
\newcommand{\kT}{\mathcal{T}}
\newcommand{\kV}{\mathcal{V}}
\newcommand{\ga}{\alpha}
\newcommand{\gb}{\beta}
\newcommand{\eps}{\varepsilon}
\newcommand{\gG}{\Gamma}
\newcommand{\gl}{\lambda}
\newcommand{\gO}{\Omega}
\newcommand{\gf}{\varphi}
\newcommand{\gF}{\Phi}
\newcommand{\gs}{\sigma}
\newcommand{\gS}{\Sigma}
\newcommand{\tm}{\subseteq}
\newtheorem{defin}{Definition}[section]
\newtheorem{prop}{Proposition}[section]
\newtheorem{theo}{Theorem}[section]
\newtheorem{lem}{Lemma}[section]
\newtheorem{cor}{Corollary}[section]
\newtheorem{rem}{Remark}[section]
\newtheorem{example}{Example}[section]
\begin{document}

\title{\Huge{Quantum Sheaves }\\ \Large{An outline of results}}

\author{
	Hans F.\ de Groote\footnote{e-mail: degroote@math.uni-frankfurt.de}
	 \\ FB Mathematik\\ J.W.Goethe-Universit\"{a}t\\ Frankfurt a.\ M.
}

\date{May 2001}
\maketitle
\bibliographystyle{plain}

\begin{abstract}
In this paper we start with the development of a theory of presheaves
on a lattice, in particular on the quantum lattice $\LL(\kH)$ of
closed subspaces of a complex Hilbert space $\kH$, and their
associated etale spaces. Even in this early state the theory has
interesting applications to the theory of operator algebras and the
foundations of quantum mechanics. Among other things we can show that
classical observables (continuous functions on a topological space)
and quantum observables (selfadjoint linear operators on a Hilbert
space) are on the same structural footing.
\end{abstract}

\section{Introduction}

Classically, an observable is a real valued (measurable, continuous or
differentiable) function on the phase space $M$ of a physical system.
In quantum mechanics, however, an observable is a selfadjoint linear
operator defined on a dense subspace of a complex Hilbert space $\kH$.

Apparently, these are quite different concepts. One of the aims of our
work is to show that both concepts are on the same structural footing.
This insight is an outcome of a theory of presheaves on an arbitrary
\emph{lattice}\footnote{For the definition of a lattice see
definition \ref{2A}. A lattice in our sense has nothing to do with
the following notion in use: a group isomorphic to a subgroup of the abelian
group $\ZZ^{d}$ for some $d \in \NN$. Here we use ``lattice'' in the
sense of the german ``Verband'', whereas the other meaning is called
in german ``Gitter''.} which we begin to study here.

The theory of presheaves and sheaves, and in particular the
cohomology theory of sheaves, is an indispensable tool in Penrose's
\emph{twistor theory} (\cite{PenRind, WardWells}). Butterfield,
Isham and Hamilton have used presheaves in the sense of topos theory
for a new interpretation of the \emph{Kochen-Specker theorem} in
quantum mechanics (\cite{ButIsh1, ButIsh2, ButIshHam}). There
are also attempts to use sheaf theory in the development of a theory
of quantum gravity(\cite{MalRap,Rap}). So sheaf theory begins to
establish in mathematical physics.

A presheaf $\kS$ on a topological space $M$ assigns to every open
subset $U$ of $M$ a set $\kS(U)$ (or a set $\kS(U)$ with some
algebraic structure: e.g. an $R$-module, a vectorspace or an algebra)
and to each pair $(U,V)$ of open sets so that $U \tm V$ a
``restriction map'' $\rho_{U}^{V} : \kS(V) \to \kS(U)$ that respects
the algebraic structure of the sets $\kS(U), \kS(V)$. A complete
presheaf (usually called a sheaf) on $M$ is a presheaf that has the
property that one can glue together compatible local data to global
ones in a unique manner (definition \ref{2F}). The definition of
presheaves and sheaves can be translated litterally from the lattice
$\kT(M)$ of open subsets of $M$ to an arbitrary lattice $\LL$. The
most important lattice that we have in mind is the orthocomplemented
lattice $\LL(\kH)$ of closed subspaces of a complex Hilbert space
$\kH$. This is the ``quantum lattice'': it is isomorphic to the
lattice of orthogonal projections of the Hilbert space $\kH$.\\
Each serious mathematical theory has to present some interesting and
convincing examples. So our first result is a disappointment: there
are no non-trivial complete presheaves on the quantum lattice
$\LL(\kH)$. But there are non-trivial presheaves on $\LL(\kH)$. There
are canonical ones, namely presheaves of spectral families in
$\LL(\kH)$, that we shall discuss in section 6.\\
In ordinary sheaf theory (over topological spaces) there is a natural
construction that assigns to each presheaf a sheaf, namely the sheaf
of local sections of the etale space of the presheaf. The etale space
of a presheaf is the space of germs of elements $f \in \kS(U) \quad (U \in
\kT(M))$
at points of $U$. We can generalize the notion of a ``point'' so that
it makes sense in an abstract (complete) lattice (definition \ref{3A}).
Unfortunately, in important lattices like $\LL(\kH)$ there are no
points at all. For the definition of germs, however, we do not need
points but only ``filter bases''. For \emph{maximal filter bases}
(these are not ultrafilters in general!) we have coined the name
\emph{quasipoints}, for they are substitutes for the non-existing
points in a general lattice. It turns out that quasipoints are already
known in lattice theory: they are the maximal dual ideals of the
lattice. The set of quasipoints of a lattice carries a natural topology.
This topology has been introduced for the case of Boolean algebras by
M.H. Stone in the thirties (of the bygone bestial century). Therefore
we call the set of quasipoints of a general lattice with its natural
topology the \emph{Stonean space} of the lattice.\\
We study Stonean spaces in section 4 and show how to associate a sheaf
of local sections over the Stonean space $\kQ(\LL)$ to a presheaf on a
lattice $\LL$.\\
In section 5 we consider the maximal distributive sublattices of the
quantum lattice $\LL(\kH)$ (called \emph{Boolean sectors}) and their
Stonean spaces. The Boolean sectors $\BB \tm \LL(\kH)$ are in one to
one correspondence to the maximal abelian von Neumann subalgebras of
the algebra $\kL(\kH)$ of bounded linear operators on $\kH$ (theorem
\ref{5D}). Moreover, if $C^{*}(\BB)$ denotes the $C^{*}$-algebra
generated by the projections $P_{U} \ \ (U \in \BB)$, then the
Gelfand spectrum $\gO_{\BB}$ of $C^{*}(\BB)$ can be identified with
the Stonean space $\kQ(\BB)$ of the Boolean sector $\BB$ (theorem
\ref{5K}).\\
In section 6 we consider a canonical example of a presheaf on
$\LL(\kH)$.This presheaf consists of spectral families of selfadjoint
operators on $\kH$, i.e. of quantum mechanical observables. We show
that one can define restriction maps in close analogy to the lattice
formulation of the restriction of functions. The restriction of
spectral families to one dimensional subspaces of $\kH$ define
functions on the projective Hilbert space $\PP(\kH)$. These functions
can be characterized abstractly without any reference to linear
operators on $\kH$ (theorem \ref{6G}). We therefore call these
functions \emph{observable functions}. They induce upper
semicontinuous functions on the Stonean space of the quantum lattice
$\LL(\kH)$ and on the Stonean spaces $\kQ(\BB)$ of each Boolean
sector $\BB \tm \LL(\kH)$. The induced observable functions on
$\kQ(\BB)$ are precisely the Gelfand transforms of selfadjoint
operators in $C^{*}(\BB)$ (theorem \ref{6N}). Using these concepts,
we show (theorem \ref{6S}) that in a precise measure theoretical sense
the number $<\rho; A> := tr(\rho A)$, where $A$ is a bounded
selfadjoint operator (an observable) and $\rho$ a positive operator of
trace $1$ (a state), is an expectation value.

In section 7 we show how continuous real valued functions on a topological
space $M$ (classical observables) can be characterized by spectral
families with values in the lattice $\kT(M)$ (theorem \ref{7K}). This
shows that classical and quantum mechanical observables are on the
same structural footing: either as functions or as spectral families.

The results presented here show that the theory of \emph{quantum
sheaves}, i.e. the theory of presheaves on a lattice and their etale
spaces, deserves to be developed further.\\
In this paper only few of the results are proved in detail. Rather
long proofs of the main results are only sketched. A detailed version
will appear in appropriate form.
\vspace{8mm}

Thanks are due especially to Andreas D\"{o}ring for several
discussions on the foundations of quantum mechanics.
\pagebreak

\section{ Preliminaries}

\begin{defin}\label{2A}
 A \emph{lattice} is a partially ordered set $(\LL, \leq)$
with a \emph{zero element} $0$ ( i.e. $\forall a \in \LL : 0\leq a$),
a \emph{unit element} $1$ (i.e. $\forall a\in \LL : a\leq 1$), such
that any two elements $a,b \in\LL$ posess a \emph{maximum} $a\vee b
\in\LL$ and a \emph{minimum} $a\wedge b \in\LL$.\\
Let $\frm$ be an infinite cardinal number.\\
The lattice $\LL$ is called $\mathfrak{m}$-complete, if every family
$(a_{i})_{i\in I}$ has a maximum $\bigvee_{i \in I}a_{i}$ and a
minimum $\bigwedge_{i\in I}a_{i}$ in $\LL$, provided that $\# I\leq
\mathfrak{m}$ holds.
A lattice $\LL$ is simply called complete, if every family
$(a_{i})_{i \in I}$ in $\LL$ (without any restriction of the
cardinality of $I$) has a maximum and a minimum in $\LL$.\\
A lattice $\LL$ is called \emph{distributive} if the two distributive
laws
\begin{eqnarray*}
	a \wedge (b \vee c) & = & (a \wedge b) \vee (a \wedge c)  \\
	a \vee (b \wedge c) & = & (a \vee b) \wedge (a \vee c)
\end{eqnarray*}
hold for all elements $a, b, c \in \LL$.
\end{defin}

$\bigvee_{i \in I}a_{i}$ is characterized by the following universal
property:
\begin{enumerate}
	\item  $\forall j\in I :\quad a_{j}\leq \bigvee_{i \in I}a_{i}$

	\item  $\forall c\in \LL :\quad ((\forall i\in I : a_{i} \leq c)
	\Rightarrow \bigvee_{i}a_{i} \leq c ).$
\end{enumerate}
An analogouos universal property characterizes the minimum
$\bigwedge_{i}a_{i}$.

Note that if  $\LL$ is a distributive complete lattice, then in
general
\begin{displaymath}
	a \wedge (\bigvee_{i \in I}b_{i}) \ne \bigvee_{i \in I}(a \wedge
	b_{i}),
\end{displaymath}
so completeness and distributivity do not imply \emph{complete
distributivity}!

Let us give some important examples.
\begin{example}\label{2B}
Let $M$ be a topological space and $\kT(M)$ the topology of $M$, i.e.
the set of all open subsets of $M$. $\kT(M)$ is a distributive
complete lattice. The maximum of a family $(U_{i})_{i \in I}$ of open
subsets $U_{i}$ of $M$ is given by
\begin{displaymath}
	\bigvee_{i \in I}U_{i} = \bigcup_{i \in I}U_{i},
\end{displaymath}
the minimum, however, is given by
\begin{displaymath}
	\bigwedge_{i \in I}U_{i} = int(\bigcap_{i \in I}U_{i}),
\end{displaymath}
where $int N$ denotes the interior of a subset $N$ of $M$.
\end{example}
\begin{example}\label{2C}
If $U \in \kT(M)$, then always
\begin{displaymath}
	U \tm int \bar{U},
\end{displaymath}
but $U \ne int \bar{U}$ in general. $U$ fails to be the interior of
its adherence $\bar{U}$, if for example $U$ has a ``crack'' or is
obtained from an open set $V$ by deleting some points of $V$.\\
We call $U$ a {\bf regular open set}, if $U = int \bar{U}$. Each $U
\in \kT(M)$ has a {\bf pseudocomplement}, defined by
\begin{displaymath}
	U^{c} := M \setminus \bar{U},
\end{displaymath}
and together with the operation of pseudocomplementation $\kT(M)$ is
a \mbox{{\bf Heyting algebra}}:
\begin{displaymath}
	\forall \ U \in \kT(M) : \ U^{ccc} = U^{c}.
\end{displaymath}
$U \in \kT(M)$ is regular if and only if $U = U^{cc}$. Let
$\kT_{r}(M)$ be the set of regular open subsets of $M$. If $U, V \in
\kT_{r}(M)$, then also $U \cap V \in \kT_{r}(M)$. The union of two
regular open sets, however, is not regular in general. Therefore one
is forced to define the maximum of two elements $U, V \in \kT_{r}(M)$
as
\begin{displaymath}
	U \vee V := (U \cup V)^{cc}.
\end{displaymath}
It is then easy to see that $\kT_{r}(M)$ is a distributive complete
lattice with the lattice operations
\begin{displaymath}
	U \wedge V := U \cap V, \quad U \vee V := (U \cup V)^{cc}.
\end{displaymath}
The pseudocomplement on $\kT(M)$, restricted to $\kT_{r}(M)$, gives
an {\bf orthocomplement} $U \mapsto U^{c}$ on $\kT_{r}(M)$:
\begin{displaymath}
	U^{cc} = U, \ U^{c} \vee U = M, \ U^{c} \wedge U = \emptyset, \ (U
	\wedge V)^{c} = U^{c} \vee V^{c}
\end{displaymath}
for all $U, V \in \kT_{r}(M)$. Thus $\kT_{r}(M)$ is a {\bf complete
Boolean lattice}.
\end{example}
\begin{example}\label{2D}
Let $M$ be a topological space and $\kB(M)$ the set of Borel subsets
of $M$. $\kB(M)$ together with the usual set theoretic operations is a
distributive $\aleph_{0}$-complete Boolean lattice, usually called
the $\gs$-algebra of Borel subsets of $M$.
\end{example}
\begin{example}\label{2E}
Let $\kH$ be a (complex) Hilbert space and $\LL(\kH)$ the set of all
closed subspaces of $\kH$. $\LL(\kH)$ is a complete lattice with
lattice operations defined by
\begin{eqnarray*}
	U \wedge V & := & U \cap V  \\
	U \vee V & := & (U + V)^{-}  \\
	U^{\perp} & := & \text{orthogonal complement of $U$ in $\kH$.}
\end{eqnarray*}
Contrary to the foregoing examples $\LL(\kH)$ is highly {\bf
non-distributive}!\\
Of course $\LL(\kH)$ is isomorphic to the lattice $P(\kL(\kH)) :=
\{P_{U} \mid U \in \LL(\kH) \}$ of all orthogonal projections in the
algebra $\kL(\kH)$ of bounded linear operators of $\kH$. The
non-distributivity of $\LL(\kH)$ is equivalent to the fact that two
projections $P_{U}, P_{V} \in P(\kL(\kH))$ do not commute in
general.\\
$\LL(\kH)$ is the basic lattice of quantum mechanics (\cite{Jauch}).
It represents the ``quantum logic'' in contrast to the classical
``Boolean logic''.
\end{example}

Traditionally, the notions of a {\bf presheaf} and a {\bf complete
presheaf} (complete presheaves are usually called ``sheaves'') are
defined for the lattice $\kT(M)$ of a topological space $M$. The very
definition of presheaves and sheaves, however, can be formulated also
for an arbitrary lattice:
\begin{defin}\label{2F}
A {\bf presheaf} of sets ($R$-modules) on a lattice $\LL$ assigns to
every element $a \in \LL$ a set ($R$-module) $\kS(a)$ and to every
pair $(a, b) \in \LL \times \LL$ with $a \leq b$ a mapping
($R$-module homomorphism)
\begin{displaymath}
	\rho_{a}^{b} : \kS(b) \to \kS(a)
\end{displaymath}
such that the following two properties hold:
\begin{enumerate}
	\item  [(1)] $\rho_{a}^{a} = id_{\kS(a)}$ for all $a \in \LL$,

	\item  [(2)] $\rho_{a}^{b} \circ \rho_{b}^{c} = \rho_{a}^{c}$ for
	all $a, b, c \in \LL$ such that $a \leq b \leq c$.
\end{enumerate}
The presheaf $(\kS(a), \rho_{a}^{b})_{a \leq b}$ is called a {\bf
complete presheaf} (or a {\bf sheaf} for short) if it has the
additional property
\begin{enumerate}
	\item  [(3)] If $a = \bigvee_{i \in I}a_{i}$ in $\LL$ and if $f_{i}
	\in \kS(a_{i}) \ \ (i \in I)$ are given such that
	\begin{displaymath}
		\forall \ i, j \in I : \ (a_{i} \wedge a_{j} \ne 0 \quad
	\Longrightarrow \quad \rho_{a_{i}\wedge a_{j}}^{a_{i}}(f_{i}) =
	\rho_{a_{i} \wedge a_{j}}^{a_{j}}(f_{j}),
	\end{displaymath}
	then there is exactly one $f \in \kS(a)$ such that
	\begin{displaymath}
		\forall \ i \in I : \ \rho_{a_{i}}^{a}(f) = f_{i}.
	\end{displaymath}
\end{enumerate}
\end{defin}
The mappings $\rho_{a}^{b} : \kS(b) \to \kS(a)$ are called {\bf
restriction maps}.

One of the most elementary and at the same time instructive examples
is the sheaf of locally defined continuous complex valued functions on
a topological space $M$: $\kS(U)$ is the space of continuous functions
on the open set $U \tm M$ and for $U, V \in \kT(M)$ with $U \tm V$
\begin{displaymath}
	\rho_{U}^{V} : \kS(V) \to \kS(U)
\end{displaymath}
is the restriction map $f \mapsto f|_{U}$. Property $(3)$ in
definition \ref{2F} expresses the elementary fact that one can glue
together a family of locally defined continuous functions $f_{i} :
U_{i} \to \CC$ which agree on the non-empty overlaps $U_{i} \cap
U_{j}$ to a continuous function $f$ on $\bigcup_{i \in I}U_{i}$ which
coincides with $f_{i}$ on $U_{i}$ for each $i \in I$.

Are there interesting new examples for sheaves on a lattice other
than $\kT(M)$, in particular on the quantum lattice $\LL(\kH)$?

The story begins with a disappointing answer:
\begin{prop}\label{2G}
Let $(\kS(U), \rho_{U}^{V})_{U \tm V}$ be a complete presheaf of sets
on the quantum lattice $\LL(\kH)$. Then
\begin{displaymath}
	\#\kS(U) = 1
\end{displaymath}
for all $U \in \LL(\kH) \setminus \{0\}$.\\
Thus complete presheaves on $\LL(\kH)$ are completely trivial!
\end{prop}
The proof of this result is based on the following observation: For
each \mbox{ $U \in \LL(\kH) \setminus \{0\}$} we have
\begin{displaymath}
	U = \bigvee_{\CC x \tm U}\CC x,
\end{displaymath}
and if $\CC x, \CC y \tm U$ are different one dimensional subspaces,
then \mbox{$\CC x \cap \CC y = 0$.} Hence the compatibility conditions in
$(3)$ of definition \ref{2F} are void and therefore
\begin{displaymath}
	\kS(U) \cong \prod_{\CC x \tm U}\kS(\CC x).
\end{displaymath}

Also on the lattice $\kT_{r}(M)$ (although a distributive complete
lattice of open sets) there are only trivial sheaves.

There are, however, non-trivial presheaves on $\LL(\kH)$ and one of
them, which we shall study in section 6, turns out to be quite
fruitful for quantum mechanics and the theory of operator algebras.

Moreover, there is also another perspective of sheaves: the etale
space of a presheaf. Classically, for a topological space $M$, a
presheaf $\kS$ on $\kT(M)$ induces a sheaf of local sections of the
etale space of $\kS$. This sheaf on $\kT(M)$ is called the
``sheafification of the presheaf $\kS$''.\\
We can imitate this natural construction in the general case of a
lattice $\LL$ and shall obtain a sheaf - not on $\LL$ because of the
foregoing result - on the lattice $\kT(\kQ(\LL))$ where $\kQ(\LL)$ is
the Stonean space of the lattice $\LL$.
\pagebreak

\section{ Points and Quasipoints in a lattice}

Let $M$ and $N$ be topological spaces. The elements of $N$ are in
one-to-one correspondence to the \emph{constant} mappings $f: M \to
N$. These constant mappings correspond via the inverse image
morphisms
\[ V \mapsto \overset{-1}{f}(V) \quad (V \in \kT(N))\]
to the continuous lattice morphisms
\[ \gF : \kT(N) \to \kT(M) \]
with the property
\[ \forall V \in \kT(N) : \gF(V) \in \{\emptyset, M \}. \]
It is immediate that the set
\[ \frp := \{ V \in \kT(N) \ |\ \gF(V) = M \} \]
has the following properties:
\begin{enumerate}
 \item[(1)] $\emptyset \notin \frp$ .

 \item[(2)] If $V,W \in \frp$, then $V\cap W \in \frp$.

 \item[(3)] If $V \in \frp$ and $W\supseteq V$ in $\kT(N)$, then $W\in
	\frp$.

 \item[(4)] If $(V_{\iota})_{\iota \in I}$ is a family in $\kT(N)$
	and $\bigcup_{\iota \in I}V_{\iota} \in \frp$, then there is at least
	one $\iota_{0} \in I$ such that $V_{\iota_{0}} \in \frp$.
\end{enumerate}
Now these properties make perfectly  sense in an arbitrary
$\frm$-complete lattice, so we can use them to define \emph{points in
a lattice}:
\begin{defin}\label{3A} Let $\LL$ be an $\frm$-complete lattice. A non-empty
subset $\frp \subseteq \LL$ is called a {\bf point in $\LL$} if the
following properties hold:
\begin{enumerate}
	\item  [(1)] $0 \notin \frp$.

	\item  [(2)] $a,b \in \frp \Rightarrow a \wedge b \in \frp$.

	\item  [(3)] $a \in \frp, b \in \LL, a \leq b \Rightarrow b \in \frp$.

	\item  [(4)] Let $(a_{\iota})_{\iota \in I}$ be a family in $\LL$
	such that $\#I  \leq \frm$ and $\bigvee_{\iota \in I}a_{\iota} \in
	\frp$ then $a_{\iota} \in \frp$ for at least one $\iota \in I$.
\end{enumerate}
\end{defin}
\begin{example}\label{3B} Let $M$ be a non-empty set and $\LL \subseteq
pot(M)$ an $\frm$-complete lattice such that
\begin{eqnarray*}
	0_{\LL} & = & \emptyset  \\
	1_{\LL} & = & M  \\
	\bigvee_{\iota \in I}U_{\iota} & = & \bigcup_{\iota \in I}U_{\iota}
	\qquad (\# I \leq \frm).
\end{eqnarray*}
Then for each $x \in M$
\[  \frp_{x} := \{ U \in \LL \ | \ x \in U \}  \]
is a point in $\LL$.
\end{example}

Conversely, if $\LL$ is the lattice $\kT(M)$ of open sets of a
regular topological space $M$ we have
\begin{prop}\label{3C} Let $M$ be a regular topological space. A
non-empty subset $\frp \subseteq \kT(M)$ is a point in the lattice
$\kT(M)$ if and only if $\frp$ is the set of open neighbourhoods of an
element $x \in M$. $x$ is uniquely determined by $\frp$.
\end{prop}

Unfortunately there are important lattices that do not possess any
points!\\
There are plenty of points in $\kT(M)$ and $\kB(M)$; $\kT_{r}(M)$
and $\LL(\kH)$ possess no points at all. We will show this
here only for the lattice $\LL(\kH)$ of closed subspaces of the
Hilbert space $\kH$.
\begin{prop}\label{3D} If $dim \kH > 1$, then there are no points in
$\LL(\kH)$.
\end{prop}
\emph{Proof:} Let $\frp \subseteq \LL(\kH)$ be a point. If
$(e_{\alpha})_{\alpha \in A}$ is an orthonormal basis of $\kH$ then
\[  \bigvee_{\alpha \in A}\CC e_{\alpha} = \kH \in \frp,  \]
so $\CC e_{\alpha_{0}} \in \frp $ for some $\alpha_{0} \in A$. It
follows that each $U \in \frp$ must contain the line $\CC
e_{\alpha_{0}}$. Now choose $U\in \LL(\kH)$ such that neither $U$ nor
$U^{\perp}$ contains $\CC e_{\alpha_{0}}$. Then $U, U^{\perp} \notin
\frp$ but $U \vee U^{\perp} = \kH \in \frp$ which is a contradiction
to property $(4)$ in the definition of a point in a lattice.\\
Therefore there are no points in $\LL(\kH)$. $\square$

Let $\kP = (\kP(U),\rho^{U} _{V})_{V\leq U}$ be a presheaf on the
topological space $M$. The \emph{stalk} of $\kP$ at $x \in M$ is the
direct limit
\[  \kP_{x} := \lim_{\overset{\longrightarrow}{U \in \frU(x)}}\kP(U) \]
where $\frU(x)$ denotes the set of open neighbourhoods of $x$ in $M$,
i.e. the point in $\kT(M)$ corresponding to $x$.\\
For the definition of the direct limit, however, we do not need the
point $\frU(x)$, but only a partially ordered set $I$ with the property
\[  \forall \ \alpha, \beta \in I\ \  \exists \gamma \in I: \gamma \leq
\alpha \ \  \mbox{and}  \ \ \gamma \leq \beta. \]
In other words: a \emph{filter basis} $B$ in a lattice $\LL$ is
sufficient. It is obvious how to define a filter basis in an
arbitrary lattice $\LL$:
\begin{defin}\label{3E} A filter basis $B$ in a lattice $\LL$ is a
non-empty subset $B \subseteq \LL$ such that
\begin{enumerate}
	\item  [(1)] $0 \notin B$,

	\item  [(2)] $\forall \ a,b \in B \ \exists \ c \in B : \ c \leq a
	\wedge b$.
\end{enumerate}
\end{defin}

The set of all filter bases in a lattice $\LL$ is of course a vast object.
So it is reasonable to consider \emph{maximal} filter bases in
$\LL$. (By Zorn's lemma, every filter basis is contained in a maximal
filter basis in $\LL$.) This leads to the following
\begin{defin}\label{3F} A nonempty subset $\frB$ of a lattice $\LL$ is
called a {\bf quasipoint} in $\LL$ iff
\begin{enumerate}
	\item  [(1)] $0 \notin \frB$

	\item  [(2)] $\forall \ a,b \in \frB \ \exists \ c \in \frB : \ c
	\leq a \wedge b$

	\item  [(3)] $\frB$ is a maximal subset having the properties $(1)$
	and $(2)$.
	\end{enumerate}
\end{defin}
\begin{prop}\label{3G} Let $\frB$ be a quasipoint in the lattice
$\LL$. Then
\[ \forall \ a \in \frB \ \forall \ b \in \LL \ :(a \leq b
\Longrightarrow b \in \frB ). \]
In particular
\[ \forall \ a, b \in \frB \ : a \wedge b \in \frB \].
\end{prop}
\emph{Proof:} Let $c \in \frB$. Then $a \wedge c \leq b \wedge c$ and
from $a, c \in \frB$ we obtain a $d \in \frB$ such that
\[  d \leq a \wedge c \leq b \wedge c. \]
Therefore $\frB \cup \{ b\}$ is a filter basis in $\LL$ containing
$\frB$. Hence $\frB = \frB \cup \{b\}$ by the maximality of $\frB$,
i.e. $b \in \frB$. $\square$

This proposition shows that a quasipoint in $\LL$ is nothing else but
a \emph{maximal dual ideal} in the lattice $\LL$ (\cite{birk}).

We shall now determine the quasipoints in some important lattices.

Let $M$ be a locally compact Hausdorff space and $\LL := \kT(M)$.
Let $\frB$ be a quasipoint in $\LL$. We distinguish two cases. In the
first case we assume that $\frB$ has an element that is a relatively
compact open subset of $M$. Let $U_{0} \in \frB$ be such an element.
Then
\[  \bigcap_{U \in \frB}\bar{U} \ne \emptyset ,  \]
for otherwise  $\bigcap_{U \in \frB}\overline{U \cap U_{0}} =
\emptyset$ and from the compactness of $\bar{U_{0}}$ we see that
there are $U_{1},\ldots,U_{n} \in \frB$ such that $\bigcap_{i =
1}^{n}\overline{U_{i} \cap U_{0}} = \emptyset$.\\ But then $U_{0} \cap
U_{1} \cap \ldots \cap U_{n} = \emptyset$, contrary to the defining
properties of a filter basis. The maximality of $\frB$ implies that
every open neighbourhood of $x \in \bigcap_{U \in \frB}\bar{U}$
belongs to $\frB$. Therefore, as $M$ is a Hausdorff space, $\bigcap_{U
\in \frB}\bar{U}$ consists of precisely one element of $M$. We will
denote this element by $pt(\frB)$ and call $\frB$ a \emph{quasipoint
over $x = pt(\frB)$}.\\
Now consider the other case in which no element of the quasipoint
$\frB$ is relatively compact. It can be easily shown, using the
maximality of $\frB$ again, that in this case $M\setminus K \in \frB$
for every \emph{compact} subset $K$ of $M$. We summarize these facts
in the following
\begin{prop}\label{3H}
Let $M$ be a locally compact Hausdorff space and $\frB$ a quasipoint
in the lattice $\kT(M)$ of open subsets of $M$. Then either
$M\setminus K \in \frB$ for all compact subsets $K$ of $M$ or there
is a unique element $x \in M$ such that $\bigcap_{U \in
\frB}\bar{U} = \{x\}$.\\
In the first case $\frB$ is called an unbounded quasipoint, in the
second case a bounded quasipoint over $x$.\\
For a non-compact space $M$ let $M_{\infty} := M \cup \{\infty\}$ be
the one-point compactification of $M$. Then the unbounded quasipoints
in $\kT(M)$ can be considered as quasipoints over $\infty$ in
$\kT(M_{\infty})$.
\end{prop}

Next we consider a \emph{Boolean $\sigma$-algebra $\kB$}, i.e. a
$\sigma$-complete complemented distributive lattice. As a consequence
of distributivity we have the following
\begin{lem}\label{3I}
Let $A \mapsto A^c$ be the complement operation in the
$\sigma$-algebra $\kB$. Then a filter basis $\frB \subseteq \kB$ is a
quasipoint in $\kB$ iff
\[  \forall \ A \in \kB \ : \ A \in \frB \ \mbox{or} \ A^c \in
\frB  .\]
Consequently every point in $\kB$ is a quasipoint in $\kB$.
\end{lem}

Moreover, we can show :
\begin{prop}\label{3J}
Let $\mathcal{I}$ be a $\sigma$-ideal in the $\sigma$-algebra $\kB$
and let $\pi : \kB \to \kB/\mathcal{I}$ be the canonical projection
onto the quotient $\sigma$-algebra $\kB/\mathcal{I}$. Then $\frB
\subseteq\kB/\mathcal{I}$ is a quasipoint if and only if
$\overset{-1}{\pi}(\frB)$ is a quasipoint in $\kB$ such that
\[ \overset{-1}{\pi}(\frB) \cap \mathcal{I} = \emptyset \].
\end{prop}
By a theorem of Loomis and Sikorski (\cite{Sik}) every
$\sigma$-algebra is the quotient of a $\sigma$-algebra of Borel sets of
a compact space modulo a $\sigma$-ideal.\\
It is not difficult to determine the quasipoints in the
$\sigma$-algebra of all Borel subsets of a topological space
(satisfying some mild topological conditions). Thus the quasipoints in
a $\sigma$-algebra are known in principal.

Our third example is the complete lattice $\LL(\kH)$ of closed
subspaces of a Hilbert space $\kH$.\\
As in the topological situation the quasipoints in $\LL(\kH)$ fall
into two different classes:
\begin{prop}\label{3K}
Let $\frB$ be a quasipoint in $\LL(\kH)$. $\frB$ contains an element
of finite dimension if and only if there is a unique line $\CC x_{0}$ in $\kH$
such that
\[ \frB = \{ U \in \LL(\kH) \ \mid \CC x_{0} \subseteq U \} . \]
$\frB$ does not contain an element of finite dimension if and only if $W \in
\frB$ for all $W\in \LL(\kH)$ of finite codimension.
\end{prop}
\emph{Proof:} Let $U_{0} \in \frB$ be finite dimensional. Then $U \cap
U_{0} \ne 0$ for all $U \in \frB$ and therefore $\{ U \cap U_{0} \mid
U \in \frB \}$ contains an element $V_{0}$ of minimal dimension.
Hence $V_{0} \subseteq U$ for all $U \in \frB$ and by the maximality
of $\frB$ $V_{0}$ must have dimension one.\\
Assume that a quasipoint $\frB$ in $\LL(\kH)$ contains every $W \in
\LL(\kH)$ of finite codimension. Let $U$ be a finite dimensional
subspace of $\kH$. Then $U^{\perp} \in \frB$ and therefore $U \notin
\frB$ because of $U \cap U^{\perp} = 0$.\\
Let $V \in \LL(\kH)$ be of finite codimension and $V \notin \frB$. Then
there is some $U \in \frB$ such that $U \cap V = 0$. Consider the
orthogonal projection
\[ P_{V^{\perp}}: \kH \to V^{\perp} \]
onto $V^{\perp}$. $U \cap V = 0$ means that the restriction of
$P_{V^{\perp}}$ to $U$ is injective. As $V^{\perp}$ is finite
dimensional, $U$ must be finite dimensional too. $\square$

Quasipoints in $\LL(\kH)$ that contain a line are called
\emph{atomic}, otherwise they are called \emph{continuous}.\\
Whereas the structure of atomic quasipoints is trivial, the set of
continuous quasipoints mirrors the whole complexity of spectral
theory of linear operators in $\kH$.
\pagebreak

\section{Stonean spaces and the etale space of a \mbox{presheaf} on a lattice}

In 1936 M.H.Stone (\cite{stone}) showed that the set $\kQ(\kB)$ of
quasipoints in a Boolean algebra $\kB$ can be given a topology such
that $\kQ(\kB)$ is a \emph{compact zero dimensional} Hausdorff space
and that the Boolean algebra $\kB$ is isomorphic to the Boolean
algebra of all \emph{closed open} subsets of $\kQ(\kB)$. A basis for
this topology is simply given by the sets
\[ \kQ_{U}(\kB) := \{ \frB \in \kQ(\kB) \mid U \in \frB \} \]
where $U$ is an arbitrary element of $\kB$.

Of course we can generalize this construction to an arbitrary
lattice $\LL$:\\
Let $\kQ(\LL)$ be the set of quasipoints in $\LL$ and for $U \in \LL$
let
\[ \kQ_{U}(\LL) := \{ \frB \in \kQ(\LL) \mid U \in \frB \}. \]
It is quite obvious from the definition of a quasipoint that
\[ \kQ_{U \wedge V}(\LL) = \kQ_{U}(\LL) \cap \kQ_{V}(\LL) \]
holds. Hence $\{ \kQ_{U}(\LL) \mid U \in \LL \}$ is a basis for a
topology on $\kQ(\LL)$. Moreover it is easy to see, using the
maximality of quasipoints, that in this topology the sets
$\kQ_{U}(\LL)$ are open and closed. Therefore the topology defined by
the basic sets $\kQ_{U}(\LL)$ is \emph{zero dimensional}. $\kQ(\LL)$
together with this topology is called the {\bf Stonean space of the
lattice $\LL$.}
\begin{rem}\label{4A}
The Stonean space $\kQ(\LL)$ of the lattice $\LL$ is a completely
regular Hausdorff space.
\end{rem}
This follows immediately from the fact that the sets $\kQ_{U}(\LL)$
are open and closed, so their characteristic functions are continuous.

In contrast to the case of Boolean algebras, Stonean spaces are not
compact in general. The situation can be even worse, as the following
important example shows:
\begin{rem}\label{4B}
Let $\kH$ be a Hilbert space of dimension greater than one. Then the
Stonean space $\kQ(\kH) := \kQ(\LL(\kH))$ is {\bf not locally
compact.}
\end{rem}
This is an easy consequence of Baire's category theorem and the
general fact that the Stonean space $\kQ(\LL_{U})$ of the principal
ideal $\LL_{U} := \{ V \in \LL \mid V \leq U \}$ of an arbitrary
lattice $\LL$ and $U \in \LL \setminus \{0\}$ is homeomorphic to
$\kQ_{U}(\LL)$.

If $\MM$ is a complete lattice isomorphic to $\LL$ via a lattice
isomorphism $\gF : \LL \to \MM$, then it is easy to see that $\gF$
induces a homeomorphism
\[  \gF_{*} : \kQ(\LL) \to \kQ(\MM) \]
of the corresponding Stonean spaces:
\[  \gF_{*}(\frB) := \{ \gF(a) \mid a \in \frB \} . \]
The opposite conclusion, however, is not true.\\
In fact we can show that the Stonean spaces $\kQ(\kT(M))$ and
$\kQ(\kT_{r}(M))$ are homeomorphic for every topological space $M$.
But in general the lattice $\kT(M)$ of open subsets of $M$ is not
isomorphic to the lattice $\kT_{r}(M)$ of regular open subsets of
$M$, because $\kT(M)$ possesses points whereas in general
$\kT_{r}(M)$ does not.

In section 2 we have seen that $\kT_{r}(M)$ is a Boolean algebra with
complement operation
\[  U \mapsto U^c \]
where $U^c := M\setminus \bar{U}$. Now it is easy to see that
\[  U \cap V = \emptyset \quad \Longrightarrow \quad U^{cc} \cap
V^{cc} = \emptyset \]
holds for all open sets $U,V \subseteq M$. From this fact we get
\begin{lem}\label{4C}
Let $M$ be a topological space and let $\frB$ be a quasipoint in
$\kT(M)$. Then
\[  \frB^r := \{ U^{cc} \mid U \in \frB \} \]
is a quasipoint in $\kT_{r}(M)$.
\end{lem}
\begin{prop}\label{4D}
The mapping
\begin{eqnarray*}
	\rho & : & \kQ(\kT(M)) \to \kQ(\kT_{r}(M))  \\
	 &  & \frB \longmapsto \frB^r
\end{eqnarray*}
is a homeomorphism of Stonean spaces.
\end{prop}
\emph{Sketch of proof:} The first thing to show is that every
quasipoint $\frR$ in $\kT_{r}(M)$ is contained in exactly one
quasipoint in $\kT(M)$. Thus $\rho$ is a bijection. Moreover
\[ U \in \frB \quad \Longleftrightarrow \quad U^{cc} \in \frB^r \]
for every quasipoint $\frB$ in $\kT(M)$. This implies
\[ \rho(\kQ_{U}(\kT(M))) = \kQ_{U^{cc}}(\kT_{r}(M)) \]
and
\[ \rho^{-1}(\kQ_{W}(\kT_{r}(M))) = \kQ_{W}(\kT(M)), \]
i.e. $\rho$ is a homeomorphism. $\square$
\begin{cor}\label{4E}
The Stonean space $\kQ(\kT(M))$ is compact.
\end{cor}
\begin{cor}\label{4F}
Let $M$ be a compact Hausdorff space and let
\[ pt : \kQ(\kT(M)) \to M \]
be the map that assigns to $\frB \in \kQ(\kT(M))$ the element
$pt(\frB) \in M$ determined by $\bigcap_{U \in \frB}\bar{U}$. Then
the quotient topology of $M$ induced by $pt$ coincides with the given
topology of $M$.
\end{cor}
This follows from the fact that $pt$ is a continuous mapping and
therefore the quotient topology is finer than the given topology. It
cannot be strictly finer because both topologies are compact and
Hausdorff.

This result gives an extreme example for the fact that the projection
onto the quotient by an equivalence relation need \emph{not} be an open
mapping: let $M$ be a \emph{connected} compact Hausdorff space. The
compactness of the Stonean space $\kQ(\kT(M)$ implies that $pt$ is a
closed mapping. If it was also an open mapping the total
disconnectedness of $\kQ(\kT(M))$ would imply that the image $M$ of
$pt$ is totally disconnected, too. As $M$ is connected, this is only
possible for the trivial case that $M$ consists of a single element.
\vspace{1cm}

In what follows we shall show that to each presheaf on a (complete)
lattice $\LL$ one can assign a sheaf on the Stonean space $\kQ(\LL)$.
The construction is quite similar to the well-known construction
called \emph{``sheafification of a presheaf''}.\\
If $\kP$ is a presheaf, say, of modules on a topological space $M$,
i.e. on the lattice $\kT(M)$, then the corresponding etale space
$\kE(\kP)$ of $\kP$ is the disjoint union of the stalks of $\kP$ at
points in $\kT(M)$:
\[  \kE(\kP) = \coprod_{x \in M}\kP_{x} \]
where
\[  \kP_{x} = \lim_{\overset{\longrightarrow}{U \in \frU}}\kP(U). \]
Now in a general lattice we need not have points. Our most important
example for this situation is the {\bf quantum lattice $\LL(\kH)$} of
closed subspaces of the Hilbert space $\kH$. However, we always have
plenty of quasipoints, and we can define the stalk of a presheaf $\kP$
on a lattice $\LL$ over a quasipoint $\frB \in \kQ(\LL)$ in the very
same manner as in the topological situation.

Let $\kP = (\kP(U),\ \rho_{V}^{U})_{V \leq U}$ be a presheaf on the
(complete) lattice $\LL$.
\begin{defin}\label{4G}
$f \in \kP(U)$ is called equivalent to $g \in \kP(V)$ at the
quasipoint $\frB \in \kQ_{U \wedge V}(\LL)$ if and only if
\[ \exists \ W \in \frB : \ W \leq U \wedge V \ \mbox{and} \
\rho_{W}^{U}(f) = \rho_{W}^{V}(g). \]
\end{defin}
If $f$ and $g$ are equivalent at the quasipoint $\frB$ we write $f
\sim_{\frB} g$.\\
It is easy to see that $\sim_{\frB}$ is an equivalence relation. The
equivalence class of $f \in \kP(U)$ at the quasipoint $\frB \in
\kQ(\LL)$ is denoted by $[f]_{\frB}$. It is called the \emph{germ of
$f$ at $\frB$}. Note that this only makes sense if $\frB \in
\kQ_{U}(\LL)$. Let $\frB \in \kQ_{U}(\LL)$. Then we obtain a canonical
mapping
\[  \rho_{\frB}^{U} : \kP(U) \to \kP_{\frB} \]
of $\kP(U)$ onto the set $\kP_{\frB}$ of germs at the quasipoint
$\frB$, defined by the composition
\[ \kP(U) \overset{i_{U}}{\hookrightarrow} \coprod_{V \in \frB}\kP(V)
\overset{\pi_{\frB}}{\to} (\coprod_{V \in \frB}\kP(V))/ \sim_{\frB}   \]
where $i_{U}$ is the canonical injection and $\pi_{\frB}$ the
canonical projection of the equivalence relation $\sim_{\frB}$.
($\kP_{\frB} := (\coprod_{V \in \frB}\kP(V))/\sim_{\frB}$ is nothing
else but the direct limit $\varinjlim_{V \in \frB}\kP(V)$
(\cite{CondeG}) and $\rho_{\frB}^{U}(f)$ is just another notation for
the germ $[f]_{\frB}$ of $f \in \kP(U)$.)

Let $\kP$ be a presheaf on the lattice $\LL$ and
\[  \kE(\kP) := \coprod_{\frB \in \kQ(\LL)}\kP_{\frB} . \]
Moreover, let
\[ \pi_{\kP} : \kE(\kP) \to \kQ(\LL)  \]
be the projection defined by
\[  \pi_{\kP}(\kP_{\frB}) := \{\frB \}. \]
We will define a toplogy on $\kE(\kP)$ such that $\pi_{\kP}$ is a local
homeomorphism.\\
For $U \in \LL$ and $f \in \kP(U)$ let
\[ \kO_{f,U} := \{ \rho_{\frB}^{U}(f) \mid \frB \in \kQ_{U}(\LL) \}. \]
It is quite easy to see that $\{ \kO_{f,U} \mid f \in \kP(U),\ U \in
\LL \}$ is a basis for a topology on $\kE(\kP)$. Together with this
topology, $\kE(\kP)$ is called the {\bf etale space of $\kP$ over
$\kQ(\LL)$.} By the very definition of this topology the projection
$\pi_{\kP}$ is a local homeomorphism, for $\kO_{f,U}$ is mapped
bijectively onto $\kQ_{U}(\LL)$.
\vspace{5mm}

If $\kP$ is a presheaf of modules or algebras, the algebraic
operations can be transferred fibrewise to the etale space
$\kE(\kP)$.\\
Addition, for example, gives a mapping from
\[ \kE(\kP) \circ \kE(\kP) := \{ (a,b) \in \kE(\kP) \times \kE(\kP)
\mid \pi_{\kP}(a) = \pi_{\kP}(b) \} \]
to $\kE(\kP)$ defined as follows:\\
Let $f \in \kP(U),\ g \in \kP(V)$ be such that
\[ a = \rho_{\pi_{\kP}(a)}^{U}(f), \ \ b = \rho_{\pi_{\kP}(b)}(g) \]
and let $ W \in \pi_{\kP}(a)$ be some element such that $W \leq U
\wedge V$. Then
\[ a + b := \rho_{\pi_{\kP}(a)}^{W}( \rho_{W}^{U}(f) +
\rho_{W}^{V}(g)) \]
is a well defined element of $\kE(\kP)$.\\
By standard techniques one can prove that the algebraic operations
 \begin{eqnarray*}
 \kE(\kP) \circ \kE(\kP)	 &\to  & \kE(\kP)    \\
 	(a,b) & \mapsto & a - b
 \end{eqnarray*}
 (and $(a,b) \mapsto ab$ if $\kP$ is a presheaf of algebras) and
 \begin{eqnarray*}
  	\kE(\kP) & \to & \kE(\kP)  \\
  	a & \mapsto & \ga a
  \end{eqnarray*}
  (scalar multiplication with $\ga$) are continuous.
  \vspace{5mm}

  From the etale space $\kE(\kP)$ over $\kQ(\LL)$ we obtain - as in
  ordinary sheaf theory -  a complete presheaf $\kP^{\kQ}$ on the
  topological space $\kQ(\LL)$ by
  \[ \kP^{\kQ}(\kV) := \gG(\kV,\ \kE(\kP)) \]
  where $\kV \subseteq \kQ(\LL)$ is an open set and $\gG(\kV,\
  \kE(\kP))$ is the set of {\bf continuous sections of $\pi_{\kP}$
  over $\kV$}, i.e. of all continuous mappings $s_{\kV} : \kV \to
  \kE(\kP)$ such that $\pi_{\kP} \circ s_{\kV} = id_{\kV}$. If $\kP$
  is a presheaf of modules, then $\gG(\kV, \kE(\kP))$ is a module, too.
  \begin{defin}\label{4H}
  The complete presheaf $\kP^{\kQ}$ on the Stonean space $\kQ(\LL)$
  is called the {\bf sheaf associated to the presheaf $\kP$ on $\LL$}.
  \end{defin}

  We will postpone the study of the general situation to later work.
  Instead we will consider a concrete presheaf on the quantum
  lattice $\LL(\kH)$ and some of its connections to quantum mechanics
  and the theory of operator algebras.
  \pagebreak

  \section{Boolean sectors and Boolean quasipoints\\ in the quantum
  lattice}
  \vspace{1cm}

  In the following $\kH$ is a fixed complex Hilbert space and
  $\kL(\kH)$ denotes the algebra of bounded linear operators of $\kH$.

  Let $A$ be a selfadjoint (not necessarily bounded) operator of
  $\kH$. The \emph{spectral theorem} states that $A$ determines a
  unique family $(P_{\gl})_{\gl \in \RR}$ of orthogonal projections
  $P_{\gl} \in \kL(\kH)$ such that
  \begin{enumerate}
  	\item [(a)] $P_{\gl} \leq P_{\mu}$ for $\gl \leq \mu$,

  	\item [(b)] $P_{\gl} = \lim_{\mu \searrow \gl}P_{\mu}$ for all
  	$\gl \in \RR$, and

  	\item [(c)] $\lim_{\gl\rightarrow -\infty}P_{\gl} = 0, \
  	\lim_{\gl \rightarrow \infty}P_{\gl} = I$,
  \end{enumerate}
  where we understand the limits with respect to the strong operator
  topology, from which $A$ can be recovered as a Riemann-Stieltjes
  integral
  \[ A = \int_{-\infty}^{\infty}\gl dP_{\gl}. \]
  Moreover, if $A$ is bounded, then $A$ commutes with an operator $B
  \in \kL(\kH)$ if and only if $B$ commutes with every spectral
  projection $P_{\gl} \quad (\gl \in \RR)$.

  Now the properties (a), (b), (c) of the family $(P_{\gl})_{\gl \in
  \RR}$ can be reformulated in terms of the closed subspaces
  \[
  \gs_{A}(\gl) := P_{\gl}\kH \in \LL(\kH)
  \]
  as follows:
  \begin{enumerate}
  	\item [(1)]  $\gs_{A}(\gl) \tm \gs_{A}(\mu)$ for $\gl \leq \mu$,

  	\item [(2)]  $\gs_{A}(\gl) = \bigcap_{\mu > \gl}\gs_{A}(\mu)$ for all
  	$\gl \in \RR$, and

  	\item [(3)]  $\bigcap_{\gl \in \RR}\gs_{A}(\gl) = 0,\ \
\bigvee_{\gl \in
  	\RR}\gs_{A}(\gl) = \kH$.
  \end{enumerate}
  \begin{defin}\label{5A}
  A mapping $\gs : \RR \to \LL(\kH)$ with the properties (1), (2),
  (3) above is called a {\bf spectral family} in the quantum
  lattice $\LL(\kH)$.
  \end{defin}
  Now let $\DD$ be a sublattice of $\LL(\kH)$. The following fact is
  well known:
  \begin{rem}\label{5B}
  A sublattice $\DD$ of $\LL(\kH)$ is distributive if and only if for
  all $U,V \in \DD$ the orthogonal projections $P_{U}, P_{V}$ onto
  $U$ and $V$ respectively, commute.
  \end{rem}
  It is obvious from Zorn's lemma that each distributive sublattice
  of $\LL(\kH)$ is contained in a maximal distributive one.
  \begin{defin}\label{5C}
  A maximal distributive sublattice $\BB$ of $\LL(\kH)$ is called a
  {\bf Boolean sector} of $\LL(\kH)$.
  \end{defin}
  In what sense this is a ``sector'' will become clear soon.

  Boolean sectors have an important interpretation in the theory of
  operator algebras.
  \begin{theo}\label{5D}
  Boolean sectors of $\LL(\kH)$ are in one-to-one correspondence with
  maximal abelian von Neumann subalgebras of $\kL(\kH)$.
  \end{theo}
  The proof relies on the facts that
  \begin{itemize}
  	\item  a von Neumann subalgebra $\kM$ of $\kL(\kH)$ is maximal
  	abelian iff $\kM = \kM'$ where
  	\[ \kM' := \{ T \in \kL(\kH) \mid \forall \ S \in \kM : \ ST = TS
  	\}; \]

  	\item  a von Neumann subalgebra $\kA$ of $\kL(\kH)$ is generated by
  	the lattice $P(\kA)$ of the projections contained in $\kA$, i.e.
  	\[ \kA = P(\kA)''. \]
  \end{itemize}
  If $\BB \tm \LL(\kH)$ is a Boolean sector and
  \[
  P(\BB) := \{ P_{U} \mid U \in \BB \}
  \]
  is the corresponding Boolean algebra of projections then
  \[
  W^{*}(\BB) := P(\BB)''
  \]
  is a maximal abelian subalgebra of $\kL(\kH)$. Conversly, if $\kM$
  is a maximal abelian subalgebra of $\kL(\kH)$, then the lattice
  $P(\kM)$ of its projections is contained in $P(\BB)$ for some
  Boolean sector $\BB$. Using the maximality of $\kM$, one shows
  that $\kM = P(\BB)''$ holds.

  It is easy to see that Boolean sectors of $\LL(\kH)$ are
  \emph{complete Boolean algebras.}
  \begin{defin}\label{5E}
  A subset $\gb \tm \LL(\kH)$ is called a {\bf Boolean quasipoint}
  in $\LL(\kH)$ iff
  \begin{enumerate}
  	\item  [(1)] $0 \ne \gb$

  	\item  [(2)] $\forall \ U,V \in \gb \ \exists \ W \in \gb : \ W \tm
  	U \cap V$

  	\item  [(3)] $\forall \ U,V \in \gb : \ P_{U}P_{V} = P_{V}P_{U}$

  	\item  [(4)] $\gb$ is a maximal set fulfilling the properties (1),
  	(2), (3).
  \end{enumerate}
  \end{defin}
  As in the case of ordinary quasipoints in $\LL(\kH)$ the defining
  properties of Boolean quasipoints imply
  \begin{enumerate}
  	\item  [\emph{(5)}] \emph{Let $\gb$ be a Boolean quasipoint in
$\LL(\kH)$ and
  	let $V \in \LL(\kH)$ be such that $P_{U}P_{V} = P_{V}P_{U}$ for
  	all $U \in \gb$ and that $W \tm V$ for some $W \in \gb$. Then $V
  	\in \gb$.}
  \end{enumerate}
  \begin{rem}\label{5F}
  Let $\BB \tm \LL(\kH)$ be a Boolean sector. A subset $\gb \tm \BB$
  is a Boolean quasipoint in $\LL(\kH)$ if and only if $\gb$ is a
  quasipoint in the Boolean algebra $\BB$.
  \end{rem}
  Obviously every Boolean quasipoint is contained in some Boolean
  sector. The term ``sector'' is motivated by the following
  \begin{prop}\label{5G}
  Each Boolean quasipoint in $\LL(\kH)$ is contained in exactly one
  Boolean sector.
  \end{prop}
  \begin{defin}\label{5H}
  We call a Boolean quasipoint (or a Boolean sector) {\bf atomic} if
  it possesses a finite dimensional element. Otherwise we speak of a
  {\bf continuous} Boolean quasipoint or sector.
  \end{defin}
  \begin{rem}\label{5I}
  An atomic Boolean quasipoint (or Boolean sector) possesses an
  element that is a line in $\kH$. If a Boolean quasipoint $\gb$ is
  contained in an atomic quasipoint in $\LL(\kH)$, then $\gb$ is
  itself atomic and hence is contained in exactly one quasipoint in
  $\LL(\kH)$.
  \end{rem}

  There are continuous sectors in $\LL(\kH)$. To see this, consider a
  hermitean operator $T$ of $\kH$ that has no eigenvalues. Let $\BB$
  be a Boolean sector that contains the spectral family of $T$. Each
  element of $\BB$ is a $T$-invariant subspace. Hence a one
  dimensional element of $\BB$ would give eigenvectors of $T$. Thus
  each Boolean sector that contains the spectral family of $T$ is
  continuous.

  The simplest Boolean sectors correspond to diagonalizable operators.
  \begin{rem}\label{5J}
  Let $b = (e_{\ga})_{\ga \in A}$ be an orthonormal basis of $\kH$.
  Then there is exactly one Boolean sector $\BB_{b}$ that includes
  $\{ \CC e_{\ga} \mid \ga \in A\}$. The elements of $\BB_{b}$ are
  the $b$-adapted elements of $\LL(\kH)$, i.e. the closed subspaces
  $U$ of $\kH$ such that
  \[
  \forall\ \ga \in A : \ e_{\ga} \in U \ \mbox{or} \ e_{\ga} \in
  U^{\perp}.
  \]
  \end{rem}
  Each orthonormal basis of $\kH$ is contained in exactly one Boolean
  sector, and two orthonormal bases of $\kH$ that are included in the
  same Boolean sector differ only by a permutation of their members.

  We have seen that the Boolean sectors of $\LL(\kH)$ correspond to
  the maximal abelian von Neumann algebras in $\kL(\kH)$. We will now
  show that also the Stonean space $\kQ(\BB)$ of a Boolean sector
  $\BB$ has an interpretation in the context of operator algebras.

  Let $\BB$ be a Boolean sector, $P(\BB)$ the corresponding Boolean
  algebra of projections and $C^{*}(\BB)$ the {\bf $C^{*}$-algebra}
  generated by $P(\BB)$, i.e. the closure of span$P(\BB)$ in the
  norm-toplogy of $\kL(\kH)$. $C^{*}(\BB)$ is an abelian
  $C^{*}$-algebra with unity and is therefore, by the Gelfand representation
  theorem, isometrically *-isomorphic to the $C^{*}$-algebra
  $C(\gO_{\BB})$ of continuous functions $\gO_{\BB} \to \CC$ on some
  compact Hausdorff space $\gO_{\BB}$. $\gO_{\BB}$ is the set of all
  multiplicative linear functionals $\tau : C^{*}(\BB) \to \CC$,
  equipped with the weak*-topology.\\
  $\gO_{\BB}$ is called the \emph{Gelfand spectrum} or the
  \emph{space of characters} of the $C^{*}$- algebra $C^{*}(\BB)$.
  \begin{theo}\label{5K}
  The Gelfand spectrum $\gO_{\BB}$ of the $C^{*}$- algebra
  $C^{*}(\BB)$ is homeomorphic to the Stonean space $\kQ(\BB)$ of all
  quasipoints in the Boolean algebra $\BB$.\\
  With respect to this homeomorphism the strongly continuous
  characters correspond to the atomic quasipoints in $\BB$.
  \end{theo}
  \emph{Sketch of proof:}
  Let $\tau \in \gO_{\BB}$. Then it is easy to see that
  \[
  \gb_{\tau} := \{\ U \in \BB \mid \tau(P_{U}) = 1 \ \}
  \]
  is a quasipoint in $\BB$.\\
  Consider $\gs, \tau \in \gO_{\BB}$. Then $\gb_{\tau} = \gb_{\gs}$
  is equivalent to
  \[
  \forall \ U \in \BB \ : \ \tau(P_{U}) = 1 \ \Longleftrightarrow \
  \gs(P_{U}) = 1.
  \]
  Because of
  \[
  im\ \tau |_{P(\BB)} = im\ \gs |_{P(\BB)} = \{0,1\}
  \]
  this implies
  \[
  \tau |_{span(P(\BB))} = \gs |_{span(P(\BB))}.
  \]
  As $\gs$ and $\tau$ are continuous, it follows that $\gs = \tau$.
  Hence the mapping
  \begin{eqnarray*}
   	\gO_{\BB} & \to & \kQ(\BB)  \\
   	\tau & \mapsto & \gb_{\tau}
   \end{eqnarray*}
   is injective.\\

   Conversely, if $\gb \in \kQ(\BB)$ is given, we define a mapping
   \[
   \tau_{\gb} : P(\BB) \to \{0,1\}
   \]
   by
   \[
   \tau_{\gb}(P_{U}) :=
   \begin{cases}
   1    &\text{if $U \in \gb$}\\
   0    &\text{otherwise}.
   \end{cases}
   \]
   The defining properties of quasipoints show that $\tau_{\gb}$ is a
   multiplicative mapping. The technical difficulty is to prove
   that $\tau_{\gb}$ can be extended to a continuous linear mapping
   $span(P(\BB)) \to \CC$.\\
   Observe that each linear combination
   \[
   T = \sum_{k = 1}^{n}a_{k}P_{u_{k}} \in span(P(\BB))
   \]
   can be written as $T = \sum_{j = 1}^{m}b_{j}P_{V_{j}}$ with
   subspaces $V_{j} \in \BB$ that are orthogonal in pairs. We call
   this an \emph{orthogonal representation of $T$}. There is a
   canonical orthogonal representation of $T = \sum_{k =
   1}^{n}a_{k}P_{U_{k}}$, namely
   \begin{align}
   \sum_{k = 1}^{n}a_{k}P_{U_{k}} &= (a_{1}+\ldots+a_{n})P_{U_{1}\cap
   \ldots \cap U_{n}} \notag \\
   &+ \sum_{i = 1}^{n}(a_{1}+\ldots
   +\Hat{a}_{i}+\ldots+a_{n})P_{U_{1}\cap\ldots\cap
   U_{i}^{\perp}\cap\ldots\cap U_{n}} \notag \\
   &+\sum_{1\leq i < j\leq
n}^{n}(a_{1}+\ldots+\Hat{a}_{i}+\ldots+\Hat{a}_{j}+\ldots+a_{n})
   P_{U_{1}\cap\ldots\cap U^{\perp}_{i}\cap\ldots\cap
   U^{\perp}_{j}\cap\ldots\cap U_{n}} \notag \\
   &+\ldots+ \sum_{i =1}^{n}a_{i}P_{U^{\perp}_{1}\cap\ldots\cap
   U_{i}\cap\ldots\cap U^{\perp}_{n}}. \notag
   \end{align}
   If $V_{1},\ldots,V_{m} \in \BB$ are orthogonal in pairs, then
   $V_{i} \in \gb$ for at most one $i \leq m$. Moreover, there exists
   a (unique) $j_{\gb} \leq m$ such that $V_{j_{\gb}} \in \gb$ if and
   only if $V_{1} \vee\ldots\vee V_{m} \in \gb$.\\
   Hence the following definition is reasonable:
   \[
   \widetilde{\tau_{\gb}}(\sum_{j = 1}^{m}b_{j}P_{V_{j}}) :=
   \begin{cases}
   b_{j_{\gb}}   &\text{if $V_{1}\vee\ldots\vee V_{m} \in \gb$}\\
   0             &\text{otherwise}.
   \end{cases}
   \]
   One proves that this definition is independent of the orthogonal
   representation of $T \in span(P(\BB))$. Using the canonical
   orthogonal representation of $T = \sum_{k = 1}^{n}a_{k}P_{U_{k}}$
   we obtain
   \[
   \widetilde{\tau_{\gb}}(T) = a_{j_{1}}+\ldots+a_{j_{s}},
   \]
   where $j_{1},\ldots,j_{s}$ are those indices, for which
   $U_{j_{1}},\ldots,U_{j_{s}}$ are elements of $\gb$. This shows the
   linearity of $\widetilde{\tau_{\gb}}$.\\
   Continuity of $\widetilde{\tau_{\gb}}$ is obvious from
   \[
   |\sum_{j = 1}^{m}b_{j}P_{V_{j}}| = \max_{j \leq m}|b_{j}|,
   \]
   where $\sum_{j = 1}^{m}b_{j}P_{V_{j}}$ is an orthogonal
   representation of $T$. Hence $\tau_{\gb}$ has a unique extension
   to a character $\widetilde{\tau_{\gb}} \in \gO_{\BB}$. By
   construction
   \[
   \gb_{\widetilde{\tau_{\gb}}} = \gb .
   \]
   The continuity of $\tau \mapsto \gb_{\tau}$ follows from
   \[
   \tau \in N_{U,\eps}(\tau_{0}) \ \Longleftrightarrow \ \gb_{\tau}
   \in \kQ_{U}(\BB)
   \]
   where $\eps \in ]0,1[, \ \tau_{0} \in \gO_{\BB}, \ U \in \BB$ such
   that $\tau_{0}(U) = 1$ and
   \[
   N_{U, \eps}(\tau_{0}) := \{\ \tau \in \gO_{\BB} \mid \
   |\tau(P_{U}) - \tau_{0}(P_{U})| < \eps \}.
   \]
   Since $\gO_{\BB}$ and $\kQ(\BB)$ are compact, $\tau \mapsto
   \gb_{\tau}$ is a homeomorphism.

   Finally, the strong limit of the monotonous net $(P_{U})_{U \in
   \gb}$ is
   \[
   \lim_{U \in \gb}P_{U} =
   \begin{cases}
   P_{\CC x}     &\text{if $\gb$ is atomic}\\
   0             &\text{otherwise,}
   \end{cases}
   \]
   and $\tau_{\gb}(P_{U}) = 1$ for all $U \in \gb$. Hence
   $\tau_{\gb}$ is strongly continuous if and only if $\gb$ is
   atomic.\\
   $\square$

   \vspace{5mm}
   In the next section we will study a canonical presheaf on the
   quantum lattice $\LL(\kH)$ and show how it determines the
   \emph{Gelfand transform}
   \[ C^{*}(\BB) \to C(\gO_{\BB}) \]
   of the abelian $C^{*}$- algebra $C^{*}(\BB)$.
   \pagebreak
   \section{Observable functions}

   We know that there is only the trivial sheaf on the quantum
   lattice $\LL(\kH)$ of closed subspaces of the Hilbert space $\kH$.
   But what about presheaves?

   An obvious example is the following one: For $U \in \LL(\kH)$ let
   $\kP(U) := \kL(U)$ be the space of bounded linear operators $U \to
   U$ and for $V \in \LL(\kH),\ V \tm U$, we define a ``restriction map''
   \[
   \rho_{V}^{U} : \kL(U) \to \kL(V)
   \]
   by
   \[ \rho_{V}^{U}(A) := P_{V}A\mid_{V}.
   \]
   Clearly these data give a presheaf on $\LL(\kH)$.

   This example looks somewhat artificial because the restriction maps
   defined above do not coincide with the usual idea of restricting a
   mapping from its domain to a smaller set. The elements of the
   stalks of this presheaf, however, have a quantum mechanical
   interpretation.
   \begin{rem}\label{6A}
   Let $A \in \kL(U)$ and let $\frB_{\CC x} \in \kQ_{U}(\kH)$ be an
   atomic quasipoint (in $\LL(U)$). Then the germ of $A$ in $\frB_{\CC
   x}$ is given by $<Ax, x>$, where $x \in S^1(\kH) \cap \CC x$.
   \end{rem}
   Namely, if $A, B \in \kL(U)$, then $A \sim_{\frB_{\CC x}} B$ if and
   only if $P_{\CC x}AP_{\CC x} = P_{\CC x}BP_{\CC x}$. Now if $x \in
   S^1(\kH)$ then
   \[
   \forall \ z \in \kH : P_{\CC x}AP_{\CC x}z = <Ax, x><z, x>x.
   \]
   Hence $P_{\CC x}AP_{\CC x} = P_{\CC x}BP_{\CC x}$ if and only if
   $<Ax, x> = <Bx, x>$.

   If $A$ is a hermitian operator and $x\in S^1(\kH)$, then $<Ax, x>$
   is interpreted as the \emph{expectation value of the observable $A$
   when the quantum mechanical system is in the pure state $\CC x$.}

   In order to obtain a more natural example of a presheaf on
   $\LL(\kH)$, we shall reformulate the operation of restricting a
   continuous function $f : U \to \RR$ to an open subset $V \tm U$ in
   the language of lattice theory.\\
   Let $M$ and $N$ be regular Hausdorff spaces. A continuous mapping
   \mbox{$f : M \to N$} induces a lattice homomorphism
   \begin{eqnarray*}
   	\Phi_{f} : \kT(N) & \to & \kT(M)  \\
   	W & \mapsto & \overset{-1}{f}(W)
   \end{eqnarray*}
   that is continuous in the following sense:
   \[
   \Phi_{f}(\bigcup_{i\in I}W_{i}) = \bigcup_{i\in I}\Phi_{f}(W_{i})
   \]
   for each family $(W_{i})_{i\in I}$ in $\kT(N)$. Conversely:

   \begin{theo}\label{6B}
   Each continuous lattice homomorphism $\Phi : \kT(N) \to \kT(M)$
   induces a unique continuous mapping $f : M \to N$ such that
   $\Phi = \Phi_{f}$.
   \end{theo}

   The proof is based on the observation that for any point $\frp$
   in $\kT(M)$ the inverse image $\overset{-1}{\Phi}(\frp)$ is a point
   in $\kT(N)$. Because the points in $\kT(M)$ correspond to the
   elements of $M$, this gives a mapping $f : M\to N$. It is then easy
   to show that $f$ has the required properties.

   Now we can describe the restriction of a continuous mapping $f : M
   \to N$ to an open subset $U$ of $M$ in the following way:
   \begin{prop}\label{6C}
   Let $f : M \to N$ be a continuous mapping between regular Hausdorff
   spaces, $\Phi_{f} : \kT(N) \to \kT(M)$ the continuous lattice
   homomorphism induced by $f$, and $U$ an open subset of $M$. Then
   \begin{eqnarray*}
    	\Phi_{f}^{U} : \kT(N) & \to & \kT(U)  \\
    	W & \mapsto & \Phi_{f}(W) \cap U
    \end{eqnarray*}
    is a continuous lattice homomorphism and the corresponding
    continuous mapping $U \to N$ is the restriction of $f$ to $U$.
    \end{prop}

    Let $\kH$ be a Hilbert space. The observables of a quantum
    mechanical system are selfadjoint operators of $\kH$. Equivalently
    we can think of observables as spectral families in $\LL(\kH)$. To
    begin with, we restrict our attention to those spectral families
    $\gs : \RR \to \LL(\kH)$ that are \emph{bounded from above} :
    \[
    \exists \ \gl_{1} \in \RR \ : \ \gs(\gl_{1}) = \kH .
    \]
    Let $U \in \LL(\kH)$ and $\gs : \RR \to \LL(\kH)$ be a spectral
    family that is bounded from above. Then it easy to see that
    \[
    \gs^{U} : \gl \mapsto \gs(\gl) \cap U
    \]
    is a spectral family in $\LL(U)$ that is bounded from above, too.
    $\gs^{U}$ is called the {\bf restriction of $\gs$ to $U$}.\\

    Let $\gS_{ba}(\kH)$ be the set of spectral families in $\LL(\kH)$
    that are bounded from above.\\
    We define for $V \tm U$ in $\LL(\kH)$ the {\bf restriction map}
    \begin{eqnarray*}
    	\rho_{U}^{V} : \gS_{ba}(U) & \to & \gS_{ba}(V)  \\
    	\gs & \mapsto & \gs^{V}.
    \end{eqnarray*}
    Clearly, $(\gS_{ba}(U), \rho_{V}^{U})_{V \tm U}$ is a presheaf
    on $\LL(\kH)$.\\
    Let $\gS_{b}(U)$ be the set of bounded spectral families in
    $\LL(U)$. These correspond to the bounded hermitian operators of
    $U$. Obviously $(\gS_{b}(U), \rho_{V}^{U})_{V \tm U}$ is a
    sub-presheaf of $(\gS_{ba}(U), \rho_{V}^{U})_{V \tm U}$.\\
    If the spectral family $\gs : \RR \to \LL(\kH)$ is {\bf not}
    bounded from above then $\gs^{U}$ may fail to be a spectral
    family in $\LL(U)$.\\
    Of course the properties
    \begin{enumerate}
    	\item  [(1)] $\gs^{U}(\gl) \tm \gs^{U}(\mu)$ for $\gl \leq \mu$

    	\item  [(2)] $\gs^{U}(\gl) = \bigcap_{\mu > \gl}\gs^{U}(\mu) $
    	for all $\gl \in \RR$

    	\item  [(3)] $\bigcap_{\gl \in \RR}\gs^{U}(\gl) = 0$
    \end{enumerate}
    hold, but
    \[
    \bigvee_{\gl \in \RR}\gs^{U}(\gl) \ne U
    \]
    in general.
    \begin{example}\label{6D}
    Let $\kH$ be a separable Hilbert space and $(e_{n})_{n \in \NN}$
    an orthonormal basis of $\kH$. Then
    \[
    \gs(\gl) := \bigvee_{n \leq \gl}\CC e_{n} \qquad (\gl \in \RR)
    \]
    defines a spectral family in $\LL(\kH)$. One can show that this
    spectral family corresponds (up to some scaling) to the Hamilton
    operator of the harmonic oscillator. Take $x \in S^1(\kH)$ such
    that
    \[
    \forall \ n \in \NN \ : \ <x, e_{n}> \ne 0.
    \]
    This means that $x \notin \gs(\gl)$ for all $\gl \in \RR$ and hence
    \[
    \gs^{\CC x}(\gl) = \gs(\gl) \cap \CC x = 0
    \]
    for all $\gl \in \RR$. Therefore
    \begin{displaymath}
    	\bigvee_{\gl \in \RR}\gs^{\CC x}(\gl) = 0 \ne \CC x.
    \end{displaymath}
    \end{example}
    \begin{rem}\label{6DD}
    Of course we can drop the requirement
    \begin{displaymath}
    	\bigvee_{\gl \in \RR}\gs(\gl) = \kH
    \end{displaymath}
    in the definition of spectral families. Then we obtain the notion
    of a {\bf generalized spectral family}. Operators that are given
    by generalized spectral families are not necessarily densely
    defined, but their domain of definition is only dense in the
    closed subspace $\bigvee_{\gl \in \RR}\gs(\gl)$ of $\kH$.
    \end{rem}

    Let us consider the restriction of a spectral family $\gs : \RR
    \to \LL(\kH)$ to a one dimensional subspace $\CC x$ more closely.
    If $\CC x \tm \gs(\gl)$ for some $\gl \in \RR$, then the hermitian
    operator corresponding to the spectral family
    \begin{displaymath}
    	\gs^{\CC x} : \RR \to \LL(\CC x)
    \end{displaymath}
    is a (real) scalar multiple $cI_{1}$ of the identity $I_{1} : \CC
    x \to \CC x$. Now $\LL(\CC x) = \{0, \CC x \}$, hence
    \begin{displaymath}
    	\gs^{\CC x}(\gl) =
    	\begin{cases}
    	0     &\text{for $\gl < c$}\\
    	\CC x &\text{for $\gl \geq c$}
    	\end{cases}
    \end{displaymath}
    and
    \begin{displaymath}
    	c = \inf \{\gl \in \RR \mid \CC x \tm \gs(\gl) \}.
    \end{displaymath}
    Using the convention
    \begin{displaymath}
    	\inf \emptyset = \infty
    \end{displaymath}
    we obtain in this way a function on the projective Hilbert space
    $\PP\kH$ with values in $\RR \cup \{\infty\}$,
    \begin{displaymath}
    	f_{\gs} : \PP\kH \to \RR \cup \{\infty\},
    \end{displaymath}
    defined by
    \begin{displaymath}
    	f_{\gs} := \inf \{\gl \in \RR \mid \CC x \tm \gs(\gl)\}.
    \end{displaymath}
    Clearly, if $\gs$ is bounded from above then $f_{\gs}$ is bounded
    from above, too. Moreover $f_{\gs}$ is a bounded function if and
    only if $\gs$ is a bounded spectral family, i.e. the corresponding
    selfadjoint operator $A_{\gs}$ is bounded.

    The canonical topology on projective Hilbert space $\PP\kH$ is the
    quotient topology defined by the projection
    \begin{eqnarray*}
    	pr : \kH \setminus\{0\} & \to & \PP\kH  \\
    	x & \mapsto & \CC x.
    \end{eqnarray*}
    This means that a subset $\mathcal{W} \tm \PP\kH$ is open if and
    only if $\overset{-1}{pr}(\mathcal{W})$ is an open subset of $\kH
    \setminus\{0\}$.\\
    The function $f_{\gs}$ has some remarkable properties:
    \begin{prop}\label{6E}
    Let $\gs : \RR \to \LL(\kH)$ be a spectral family and let
    \begin{displaymath}
    	f_{\gs} : \PP\kH \to \RR \cup \{\infty\}
    \end{displaymath}
    be the function defined by
    \begin{displaymath}
    	f_{\gs}(\CC x) := \inf \{\gl \in \RR \mid \CC x \tm \gs(\gl) \}.
    \end{displaymath}
    Then
    \begin{enumerate}
    	\item  [(1)] $f_{\gs}$ is lower semicontinuous on $\PP\kH$;

    	\item  [(2)] if $\CC x, \CC y, \CC z$ are elements of $\PP\kH$
    	such that $\CC z \tm \CC x + \CC y$, then
    	\begin{displaymath}
    		f_{\gs}(\CC z) \leq \max (f_{\gs}(\CC x), f_{\gs}(\CC y));
    	\end{displaymath}

    	\item  [(3)] $\overset{-1}{f_{\gs}}(\RR)$ is dense in $\PP\kH$.
    \end{enumerate}
    \end{prop}
    Lower semicontinuity follows from
    \begin{displaymath}
    	\overset{-1}{pr}(\overset{-1}{f_{\gs}}(]- \infty, \gl])) \cup
    	\{0\} = \gs(\gl);
    \end{displaymath}
    for then $\overset{-1}{f_{\gs}}(]-\infty, \gl])$ is closed in
    $\PP\kH$ for all $\gl \in \RR$ and therefore $f_{\gs}$ is lower
    semicontinuous. The two other properties are obvious from the
    definitions.
    \begin{defin}\label{6F}
    A function $f : \PP\kH \to \RR \cup \{\infty\}$ is called an
    {\bf observable function} if it has the following properties:
    \begin{enumerate}
    	\item  [(1)] $f$ is lower semicontinuous;

    	\item  [(2)] if $\CC x, \CC y, \CC z$ are elements of $\PP\kH$
    	such that $\CC z \tm \CC x + \CC y$ then
    	\begin{displaymath}
    		f(\CC z) \leq \max (f(\CC x), f(\CC y));
    	\end{displaymath}

    	\item  [(3)] $\overset{-1}{f}(\RR)$ is dense in $\PP\kH$.
    \end{enumerate}
    \end{defin}
    The  point is that we can reconstruct spectral families
    in $\LL(\kH)$ from observable functions on $\PP\kH$:
    \begin{theo}\label{6G}
    The mapping $\gs \mapsto f_{\gs}$ is a bijection from the set of
    spectral families in $\LL(\kH)$ onto the set of observable
    functions on $\PP\kH$. This mapping is compatible with restrictions:
    \begin{displaymath}
    	f_{\gs^{U}} = f_{\gs}|_{\PP U}.
    \end{displaymath}
    Moreover, $\gs \in \gS_{ba}(\kH)$ if and only if
    $\overset{-1}{f_{\gs}}(\RR) = \PP\kH$, and $\gs \in \gS_{b}(\kH)$
    if and only if $f_{\gs}$ is bounded.
    \end{theo}
    \emph{Sketch of proof:} The construction of a spectral family from
    an observable function $f$ is roughly as follows: for $\gl \in
    \RR$ let
    \begin{displaymath}
    	\gs(\gl) := \overset{-1}{pr}(\overset{-1}{f}(]-\infty, \gl]))
    	\cup \{0\}.
    \end{displaymath}
    Property $(1)$ assures that $\gs(\gl)$ is closed in $\kH$ and
    property $(2)$ implies that $\gs(\gl)$ is a subspace of $\kH$. It
    is not difficult to show that $\gs : \gl \mapsto \gs(\gl)$ is a
    spectral family in $\LL(\kH)$ and that
    \begin{displaymath}
    	f_{\gs} = f
    \end{displaymath}
    holds. It follows from Baire's category theorem that $\gs \in
    \gS_{ba}(\kH)$ if and only if $\overset{-1}{f_{\gs}}(\RR) =
    \PP\kH$. \ $\square$

    If $A$ is the selfadjoint operator corresponding to the spectral
    family $\gs$, then we also write $f_{A}$ instead of $f_{\gs}$.

    The {\bf spectrum} $spec(A)$ of a selfadjoint operator on $\kH$ is
    given by the corresponding observable function $f_{A}$ in a
    surprisingly simple manner:
    \begin{prop}\label{6H}
    Let $A$ be a selfadjoint operator on $\kH$. Then
    \begin{displaymath}
    	spec(A) = \overline{f_{A}(\overset{-1}{f_{A}}(\RR))},
    \end{displaymath}
    which simplifies to
    \begin{displaymath}
    	spec(A) = \overline{f_{A}(\PP\kH)}
    \end{displaymath}
    if $A$ is bounded from above.
    \end{prop}
    In what follows we investigate the r\^{o}le of observable
    functions for the etale space corresponding to the presheaf
    $(\gS_{ba}(U), \rho_{V}^{U})_{V \tm U}$ and for the Gelfand
    representation of the $C^{*}$-algebra $C^{*}(\BB)$ of a Boolean
    sector $\BB \tm \LL(\kH)$.

    We begin with a reformulation of the definition of observable
    functions.\\
    Let $\kO(\kH)$ be the set of observable functions on $\PP\kH$,
    $\kO_{ba}(\kH)$ the set of observable functions that are bounded
    from above, and $\kO_{b}(\kH)$ the set of bounded observable
    functions.\\
    Let $f \in \kO(\kH), \ \CC x \in \overset{-1}{f}(\RR) \quad
    \text{and} \quad
    \frB_{\CC x} \in \kQ(\LL(\kH))$ the atomic quasipoint defined by
    $\CC x$. Let further $\gs$ be the spectral family corresponding
    to $f$. Then
    \begin{eqnarray*}
    	f(\CC x)   & = & \inf \{\gl \in \RR \mid \CC x \tm \gs(\gl) \}  \\
    	 & = & \inf \{\gl \in \RR \mid \gs(\gl) \in \frB_{\CC x} \}.
    \end{eqnarray*}
    Using this formulation, we can extend the definition of observable
    functions to arbitrary quasipoints in $\LL(\kH)$:
    \begin{defin}\label{6I}
    Let $f \in \kO(\kH)$ and let $\gs_{f} : \RR \to \LL(\kH)$ be the
    spectral family corresponding to $f$. The function
    \begin{displaymath}
    	\hat{f} : \kQ(\LL(\kH)) \to \RR \cup \{\infty\},
    \end{displaymath}
    defined by
    \begin{displaymath}
    	\hat{f}(\frB) := \inf \{\gl \in \RR \mid \gs(\gl) \in \frB \},
    \end{displaymath}
    is called the {\bf observable function on $\kQ(\LL(\kH))$ induced
    by $f$.}
    \end{defin}
    The observable function $\hat{f}$ induced by $f \in \kO_{b}(\kH)$
    can also be expressed directly in terms of $f$:
    \begin{prop}\label{6J}
    Let $f$ be a bounded observable function. Then the observable
    function $\hat{f}$ induced by $f$ is given by
    \begin{displaymath}
    	\forall \ \frB \in \kQ(\LL(\kH)) \ : \ \hat{f}(\frB) = \inf_{U
    	\in \frB}\sup_{\CC x \tm U}f(\CC x).
    \end{displaymath}
    \end{prop}
    \begin{prop}\label{6K}
    Let $f \in \kO_{ba}(\kH)$. Then the induced observable function
    $\hat{f} : \kQ(\LL(\kH)) \to \RR$ is bounded from above and {\bf
    upper semicontinuous}.
    \end{prop}
    From now on we will denote the observable function $\kQ(\LL(\kH))
    \to \RR \cup \{\infty\}$ induced by $f \in \kO(\kH)$ also with
    the letter $f$.

    Next we will show how observable functions $f : \kQ(\LL(\kH)) \to
    \RR$ can be used to assign a value to the germ $[\gs]_{\frB}$ of
    a spectral family $\gs$ in the quasipoint $\frB \in \kQ(\LL(\kH))$.
    We recall that spectral families $\gs \in \gS_{ba}(U)$ and $\tau
    \in \gS_{ba}(V)$ are equivalent at the quasipoint $\frB \in \kQ_{U
    \cap V}(\LL(\kH))$ if and only if there is an element $W \in \frB$
    such that $W \tm U \cap V$ and $\gs^{W} = \tau^{W}$ holds.
    \begin{prop}\label{6L}
    Let $\gs \in \gS_{ba}(U), \ \tau \in \gS_{ba}(V)$ be spectral
    families with corresponding observable functions $f_{\gs}$ and
    $f_{\tau}$ respectively. If $\gs$ and $\tau$ are equivalent at
    $\frB \in \kQ_{U \cap V}(\LL(\kH))$, then
    \begin{displaymath}
    	f_{\gs}(\frB) = f_{\tau}(\frB)
    \end{displaymath}
    holds.
    \end{prop}
    This follows directly from the observation that the definition of
    equivalence at $\frB$ implies
    \begin{displaymath}
    	\{\gl \in \RR \mid \gs(\gl) \in \frB \} = \{\gl \in \RR \mid
    	\tau(\gl) \in \frB \}.
    \end{displaymath}
    The proposition shows that we obtain a mapping
    \[	v : \kE(\gS_{ba})  \to  \RR   \]
    defined by
    \begin{displaymath}
    	    	v([\gs]_{\frB}) = f_{\gs}(\frB)
    \end{displaymath}
    on the etale space $\kE(\gS_{ba})$. $v([\gs]_{\frB})$ is called
    the {\bf value of the germ $[\gs]_{\frB}$.}

    Let us consider a simple example. Let $U$ be a non-zero element
    of $\LL(\kH)$. Then the spectral family of the orthogonal
    projection $P_{U}$ onto $U$ is given by
    \begin{displaymath}
    	\gs(\gl) =
    	\begin{cases}
    	0     &\text{for $\gl < 0$}\\
    	U^{\perp}     &\text{for $0 \leq \gl < 1$}\\
    	\kH     &\text{for $1 \leq \gl$}
    	\end{cases}
    \end{displaymath}
    and therefore the corresponding observable function $f_{\gs}$ on
    $\kQ(\LL(\kH))$ is given by
    \begin{displaymath}
    	f_{\gs}(\frB) =
    	\begin{cases}
    	0   &\text{if $U^{\perp} \in \frB$}\\
    	1   &\text{if $U^{\perp} \ne \frB$}.
    	\end{cases}
    \end{displaymath}
    Of course $U \in \frB$ implies $U^{\perp} \ne \frB$. The converse,
    however, is not true. But for a {\bf Boolean quasipoint} $\gb$ we
    have
    \begin{displaymath}
    	U^{\perp} \notin \gb \Longleftrightarrow U \in \gb.
    \end{displaymath}
    So the situation becomes much simpler for Boolean quasipoints.
    \begin{defin}\label{6M}
    Let $f : \PP\kH \to \RR$ be an observable function, $\BB \tm
    \LL(\kH)$ a Boolean sector and $\gb \in \kQ(\BB)$. Then the
    function
    \begin{displaymath}
    	f^{\BB} : \kQ(\BB) \to \RR
    \end{displaymath}
    defined by
    \begin{displaymath}
    	f^{\BB}(\gb) := \inf \{\gl \in \RR \mid \gs_{f}(\gl) \in \gb \}
    \end{displaymath}
    is called the $\BB$-observable function induced by $f$.
    \end{defin}
    We therefore obtain for $\gs = \gs_{P_{U}}$:
    \begin{displaymath}
    	f_{\gs}^{\BB} = \chi_{\kQ_{U}(\BB)}
    \end{displaymath}
    where $\chi_{\kQ_{U}(\BB)}$ denotes the characteristic function
    of the subset $\kQ_{U}(\BB) \tm \kQ(\BB)$. Now $\kQ_{U}(\BB)$ is
    open and closed in the Stonean topology of $\kQ(\BB)$, hence
    $f_{\gs}^{\BB}$ is a continuous function.\\
    This is no accident.\\
    If $\BB$ is a Boolean sector and $C^{*}(\BB)$ the $C^{*}$-algebra
    generated by $\{P_{U} \mid U \in \BB \}$, we have seen in theorem \ref{5K}
    that $\kQ(\BB)$ is homeomorphic to the Gelfand spectrum of
    $C^{*}(\BB)$. Let $C^{*}(\BB)_{sa}$ be the real subalgebra of hermitian
    elements of $C^{*}(\BB)$.\\
    \emph{We can show that $f_{A}^{\BB} \ \ (A \in C^{*}(\BB)_{sa})$ is
    a continuous function on $\kQ(\BB)$ and that the mapping
    \begin{eqnarray*}
    	C^{*}(\BB)_{sa} & \to & C(\kQ(\BB))  \\
    	A & \mapsto & f_{A}^{\BB}
    \end{eqnarray*}
    is the restriction of the Gelfand transform
    \begin{eqnarray*}
    	C^{*}(\BB) & \to & C(\kQ(\BB))  \\
    	A & \mapsto & (\hat{A} : \gb \mapsto \tau_{\gb}(A))
    \end{eqnarray*}
    to the real subalgebra $C^{*}(\BB)_{sa}$.}

    The proof proceeds in several steps. We denote by $\kA(\BB)$ the
    complex algebra generated by $\{P_{U} \mid U \in \BB \}$. In our
    special situation this is just $span_{\CC}\{P_{U} \mid U \in \BB
    \}$. $\kA(\BB)$ is dense in $C^{*}(\BB)$. Let $A \in \kA(\BB)$. We
    have called
    \begin{displaymath}
    	A = \sum_{j = 1}^{m}b_{j}P_{V_{j}}
    \end{displaymath}
    an orthogonal representation of $A$ if $V_{1},\ldots,V_{m} \in
    \BB$ are orthogonal in pairs. In the previous section we have seen
    that each element of $\kA(\BB)$ possesses at least one orthogonal
    representation. Let $\sum_{j = 1}^{m}b_{j}P_{V_{j}}$ and
    $\sum_{k = 1}^{n}c_{k}P_{W_{k}}$ be orthogonal representations of
    $A \in \kA(\BB)$. One can show that
    \begin{displaymath}
    	\sum_{j = 1}^{m}b_{j}\chi_{\kQ_{V_{j}}(\BB)} = \sum_{k =
    	1}^{n}c_{k}\chi_{\kQ_{W_{k}}(\BB)}.
    \end{displaymath}
    Therefore we obtain a mapping
    \begin{displaymath}
    	\kF_{\BB} : \kA(\BB) \to C(\kQ(\BB))
    \end{displaymath}
    defined by
    \begin{displaymath}
    	\kF_{\BB}(A) := \sum_{j}b_{j}\chi_{\kQ_{V_{j}}(\BB)}
    \end{displaymath}
    where $\sum_{j}b_{j}P_{V_{j}}$ is any orthogonal representation
    of $A$. One shows that $\kF_{\BB}$ is an \emph{isometric
    homomorphism of complex algebras.} By the Stone-Weierstrass
    theorem the subalgebra $span_{\CC}\{\chi_{\kQ_{U}(\BB)} \mid U
    \in \BB \}$ is uniformly dense in $C(\kQ(\BB))$. Therefore
    $\kF_{\BB}$ can be extended uniquely to an isometric isomorphism
    $C^{*}(\BB) \to C(\kQ(\BB))$ of $C^{*}$-algebras. We denote this
    isomorphism by $\kF_{\BB}$ again.\\
    In the next step one shows that $\kF_{\BB} : C^{*}(\BB) \to
    C(\kQ(\BB))$ coincides with the Gelfand transform of $C^{*}(\BB)$
    (where we have identified $\kQ(\BB)$ with the Gelfand spectrum
    $\gO_{\BB}$ of $C^{*}(\BB)$ according to theorem \ref{5K}).

    The second major part of the proof is to show that for $A \in
    \kA(\BB)_{sa}$, i.e. $A \in span_{\RR}\{P_{U} \mid U \in \BB \}$,
    the induced observable function
    \begin{displaymath}
    	f_{A} : \kQ(\BB) \to \RR
    \end{displaymath}
    equals $\kF_{\BB}(A)$.\\

    Finally, in the third part of the proof, one shows that the
    equality \mbox{$f_{A} = \kF_{\BB}(A)$} holds on $C^{*}(\BB)_{sa}$.

    Summarizing we obtain
    \begin{theo}\label{6N}
    Let $\BB$ be a Boolean sector of $\LL(\kH)$. Then the induced
    observable functions $f_{A}^{\BB} : \kQ(\BB) \to \RR$ are
    continuous for all $A \in C^{*}(\BB)_{sa}$ and the mapping
    \begin{eqnarray*}
    	C^{*}(\BB)_{sa} & \to & C(\kQ(\BB))  \\
    	A & \mapsto & f_{A}^{\BB}
    \end{eqnarray*}
    is the restriction of the Gelfand transform $\kF_{\BB} :
    C^{*}(\BB) \to C(\kQ(\BB))$ to $C^{*}(\BB)_{sa}$.
    \end{theo}
    The induced observable functions $f_{A} :\kQ(\LL(\kH)) \to \RR$
    and $f_{A}^{\BB} : \kQ(\BB) \to \RR$ of an hermitian operator $A$
    are closely related:
    \begin{rem}\label{6O}
    Let $\gb \in \kQ(\BB)$ and let $\frB \in \kQ(\LL(\kH))$ be a
    quasipoint that contains $\gb$. Then
    \begin{displaymath}
    	f_{A}^{\BB}(\gb) = f_{A}(\frB)
    \end{displaymath}
    for all $A \in W^{*}(\BB)_{sa}$.
    \end{rem}
    If $X$ is a topological space, we denote by $\frN(X)$ the set of
    upper semicontinuous functions $X \to \RR$.
    \begin{prop}\label{6P}
    The mapping
    \begin{eqnarray*}
    	\kL(\kH)_{sa} & \to & \frN(\kQ(\LL(\kH)))  \\
    	A & \mapsto & f_{A}
    \end{eqnarray*}
    is injective. Moreover, if $\BB \tm \LL(\kH)$ is a Boolean sector,
    the mapping
    \begin{eqnarray*}
    	W^{*}(\BB)_{sa} & \to & \frN(\kQ(\BB))  \\
    	A & \mapsto & f_{A}^{\BB}
    \end{eqnarray*}
    is injective, too.
    \end{prop}
    \begin{cor}\label{6Q}
    The hermitian operators $A \in W^{*}(\BB)$ whose induced
    observable functions $f_{A}^{\BB}$ are continuous are precisely the
    hermitian elements of $C^{*}(\BB)$.
    \end{cor}

    In the following we will show that a positive operator of finite
    trace induces a bounded positive Radon measure $\mu_{\rho}^{\BB}$
    on the compact Stonean space $\kQ(\BB)$ of each Boolean sector
    $\BB \tm \LL(\kH)$. $\mu_{\rho}^{\BB}$ will be a probability
    measure on $\kQ(\BB)$ if and only if $tr\rho = 1$.

    Let $\gf \in C(\kQ(\BB))$ and let
    \begin{displaymath}
    	\gf = \gf_{1} + i\gf_{2}
    \end{displaymath}
    be the decomposition of $\gf$ into real and imaginary part. There
    are uniquely determined hermitian operators $A_{1}, A_{2} \in
    C^{*}(\BB)_{sa}$ such that
    \begin{displaymath}
    	\gf_{k} = f_{A_{k}}^{\BB} \qquad (k = 1, 2).
    \end{displaymath}
    Then
    \begin{displaymath}
    	A_{\gf} := A_{1} + iA_{2} \in C^{*}(\BB)
    \end{displaymath}
    and we define
    \begin{displaymath}
    	f_{A_{\gf}}^{\BB} := f_{A_{1}}^{\BB} + if_{A_{2}}^{\BB}.
    \end{displaymath}
    $f_{A_{\gf}}^{\BB}$ is called the observable function
    corresponding to $A_{\gf}$. Obviously
    \[ f_{A_{\gf}}^{\BB} = \gf. \]
    \begin{prop}\label{6R}
    Let $\rho \in \kL(\kH)$ be a positive operator of finite trace.
    Then
    \begin{eqnarray*}
    	\mu_{\rho}^{\BB} : C(\kQ(\BB)) & \to & \CC  \\
    	\gf & \mapsto & tr(\rho A_{\gf})
    \end{eqnarray*}
    is a bounded positive Radon measure on the Stonean space
    $\kQ(\BB)$ of the Boolean sector $\BB \tm \LL(\kH)$. Moreover,
    $\mu_{\rho}^{\BB}$ has norm $tr\rho$, so $\mu_{\rho}^{\BB}$ is a
    probability measure on $\kQ(\BB)$ if and only if $tr\rho = 1$.
    \end{prop}
    The following result shows that the value $<Ax, x> \in \RR$ for a
    hermitian operator $A$ (an {\bf observable}) and an $x \in
    S^{1}(\kH)$ (a {\bf pure state}) is really an ``expectation
    value'' unless $x$ is an eigenvector for $A$. In that case
    the system in state $x$ answers to the observable $A$ with the
    eigenvalue $\gl$ corresponding to the eigenvector $x$.\\
    This result supports the conventional wisdom in quantum mechanics.
    \begin{theo}\label{6S}
    Let $\kH$ be a separable Hilbert space, $\BB \tm \LL(\kH)$ a
    Boolean sector and $\rho$ a state, i.e. a positive operator on
    $\kH$ of trace $1$. Then the Radon measure $\mu_{\rho}^{\BB}$
    on $\kQ(\BB)$ is the point measure $\eps_{\gb_{0}}$ for some
    $\gb_{0} \in \kQ(\BB)$, if and only if there is an $x \in
    S^1(\kH)$ such that $\CC x \in \BB, \ \gb_{0} = \gb_{\CC x}$
    and $\rho = P_{\CC x}$.
    \end{theo}
    \emph{Proof:} Let $x \in S^1(\kH)$ such that $\CC x \in \BB$ and
    let $\rho := P_{\CC x}$. Let $\gf$ be a real valued function on
    $\kQ(\BB)$ and let $A_{\gf} \in \kL(\kH)$ be the corresponding
    hermitian operator:
    \begin{displaymath}
    	\gf = f_{A_{\gf}}^{\BB}.
    \end{displaymath}
    $P_{\CC x}$ commutes with $A_{\gf}$, so $x$ is an eigenvector of
    $A_{\gf}$. Let $\gl$ be the corresponding eigenvalue. Then
    \begin{displaymath}
    	tr(\rho A_{\gf}) = <A_{\gf}x, x> = \gl<x, x> = \gl.
    \end{displaymath}
    On the other hand
    \begin{displaymath}
    	\gl = f_{A_{\gf}}(\CC x) = f_{A_{\gf}}^{\BB}(\gb_{\CC x}) =
    	\eps_{\gb_{\CC x}}(f_{A_{\gf}}^{\BB}) = \eps_{\gb_{\CC x}}(\gf).
    \end{displaymath}
    Therefore
    \begin{displaymath}
    	\mu_{\rho}^{\BB} = \eps_{\gb_{\CC x}}.
    \end{displaymath}

    Conversely, let $\rho$ be a state and $\BB$ a Boolean sector such
    that $\mu_{\rho}^{\BB}$ is the point measure $\eps_{\gb_{0}}$ for
    some $\gb_{0} \in \kQ(\BB)$. Then for all $U \in \BB$
    \begin{displaymath}
    	tr(\rho P_{U}) = \mu_{\rho}^{\BB}(\chi_{\kQ_{U}(\BB)}) =
    	\chi_{\kQ_{U}(\BB)}(\gb_{0})
    \end{displaymath}
    and therefore
    \begin{displaymath}
    	\forall \ U \in \BB : \ (U \in \gb_{0} \ \Longleftrightarrow \
    	tr(\rho P_{U}) = 1).
    \end{displaymath}
    Let $U \in \gb_{0}$ and let $(e_{k})_{k \in \NN}$ be an
    $U$-adapted orthonormal basis of $\kH$, i.e. $e_{k} \in U \cup
    U^{\perp}$ for all $k \in \NN$. Then
    \begin{eqnarray*}
    	1 & = & tr(\rho P_{U})  \\
    	 & = & tr(P_{U}\rho)  \\
    	 & = & \sum_{k}<P_{U}\rho e_{k}, e_{k}>  \\
    	 & = & \sum_{k}<\rho e_{k}, P_{U}e_{k}>  \\
    	 & = & \sum_{e_{k} \in U}<\rho e_{k}, e_{k}>.
    \end{eqnarray*}
    Because of $<\rho e_{k}, e_{k}> \geq 0$ for all $k \in \NN$ and
    $tr\rho = 1$ we conclude that
    \begin{displaymath}
    	\forall \ e_{k} \in U^{\perp} : \ <\rho e_{k}, e_{k}> = 0.
    \end{displaymath}
    Hence $\rho e_{k} = 0$ for all $e_{k} \in U^{\perp}$ and therefore
    \begin{displaymath}
    	\rho(I - P_{U}) = 0
    \end{displaymath}
    i.e.
    \begin{displaymath}
    	\rho = \rho P_{U}.
    \end{displaymath}
    In particular
    \begin{displaymath}
    	\rho P_{U} = \rho = \rho^{*} = P_{U}\rho.
    \end{displaymath}
    Now for all $y \in \kH$
    \begin{displaymath}
    	\rho y = \rho P_{U}y =P_{U}\rho y
    \end{displaymath}
    and therefore
    \begin{displaymath}
    	\forall \ U \in \gb_{0} \ \forall \ y \in \kH : \ \rho y \in U.
    \end{displaymath}
    This implies
    \begin{displaymath}
    	\forall \ U \in \gb_{0} : \ \overline{im \rho} \tm U,
    \end{displaymath}
    and from the maximality of $\gb_{0}$ we conclude that
    $\overline{im \rho} \in \gb_{0}$. Hence $\overline{im \rho} = im
    \rho = \CC x$ for a unique $\CC x \in \LL(\kH)$. $\gb_{0}$ is
    contained in exactly one Boolean sector and therefore $\CC x \in
    \BB$.\\
    There is a unique $\gl_{0} \in \CC$ such that
    \begin{displaymath}
    	\rho x = \gl_{0}x.
    \end{displaymath}
    $\rho \geq 0$ and $tr\rho = 1$ imply $\gl_{0} = 1$. Hence for all
    $y \in \kH$
    \begin{displaymath}
    	\rho^2 y = \rho(\rho y) = \rho(\gl x) = \gl\rho x = \gl x = \rho y
    \end{displaymath}
    and therefore
    \begin{displaymath}
    	\rho = P_{\CC x}. \qquad \square
    \end{displaymath}
    \pagebreak

    \section{Classical observables and spectral families}

    In the previous section we have seen that a bounded hermitian
    operator $A$ on a Hilbert space $\kH$ induces bounded upper
    semicontinuous real valued functions on the Stonean spaces
    $\kQ(\BB)$ of those Boolean sectors $\BB \tm \LL(\kH)$ that
    contain the spectral family of $A$.\\
    These functions are continuous if $A$ belongs to the adherence of
    the linear span of its spectral projections with respect to the
    norm topology of $\kL(\kH)$.

    In this section we will show that a continuous real valued
    function on a topological space $M$ can be described by a
    spectral family with values in the complete lattice $\kT(M)$ of
    open subsets of $M$. These spectral families $\gs : \RR \to
    \kT(M)$ can be characterized abstractly by a certain property of
    the mapping $\gs$. Thus also a classical observable has a
    ``quantum mechanical'' description. This shows that classical and
    quantum mechanical observables are on the same structural
    footing: either as functions or as spectral families.

    It is quite natural to generalize the definition of a spectral
    family in the quantum lattice $\LL(\kH)$ to a general complete
    lattice $\LL$:
    \begin{defin}\label{7A}
    Let $\LL$ be a complete lattice. A spectral family in $\LL$ is a
    mapping $\gs : \RR \to \LL$ with the following properties:
    \begin{enumerate}
    	\item  [(1)] $\gs(\gl) \leq \gs(\mu)$ for $\gl \leq \mu$,

    	\item  [(2)] $\gs(\gl) = \bigwedge_{\mu > \gl}\gs(\mu)$ for all $\gl
    	\in \RR$, and

    	\item  [(3)] $\bigwedge_{\gl \in \RR}\gs(\gl) = 0, \
    	\bigvee_{\gl \in \RR}\gs(\gl) = 1$.
    \end{enumerate}
    \end{defin}
    We postpone the investigation of spectral families in a general
    complete lattice to later work and concentrate here mainly on
    spectral families in the lattice $\kT(M)$ for a topological
    space $M$.

    Recall that in the lattice $\kT(M)$ the (infinite) lattice
    operations are given by
    \begin{displaymath}
    	\bigvee_{\ga \in A}U_{\ga} = \bigcup_{\ga \in A}U_{\ga}
    \end{displaymath}
    and
    \begin{displaymath}
    	\bigwedge_{\ga \in A}U_{\ga} = int(\bigcap_{\ga \in A}U_{\ga}),
    \end{displaymath}
    where $int N$ denotes the interior of the subset $N$ of $M$.\\
    We would like to begin with some simple examples:
    \begin{example}\label{7B}
    The following settings define spectral families $\gs_{id},
    \gs_{abs}, \gs_{ln}, \gs_{step}$ in $\kT(\RR)$:
    \begin{eqnarray}
    	\gs_{id}(\gl) & := & ]-\infty, \gl[,
    	\label{1}  \\
    	\gs_{abs}(\gl) & := & ]-\gl, \gl[
    	\label{2}  \\
    	\gs_{ln}(\gl) & := & ]-\exp(\gl), \exp(\gl)[
    	\label{3}  \\
    	\gs_{step}(\gl) & := & ]-\infty, \lfloor\gl\rfloor[
    	\label{4}
    \end{eqnarray}
    where $\lfloor\gl\rfloor$ denotes the ``floor of $\gl \in \RR$'':
    \begin{displaymath}
    	\lfloor\gl\rfloor = \max \{n \in \ZZ \mid n \leq \gl \}.
    \end{displaymath}
    \end{example}
    The names of these spectral families sound somewhat crazy at the
    moment, but we will justify them soon.

    In close analogy to the case of spectral families in the
    lattice $\LL(\kH)$, each spectral family in $\kT(M)$ induces a
    function on a subset of $M$.
    \begin{defin}\label{7C}
    Let $\gs : \RR \to \kT(M)$ be a spectral family in $\kT(M)$. Then
    \begin{displaymath}
    	\kD(\gs) := \{x \in M \mid \exists \ \gl \in \RR : \ x \notin
    	\gs(\gl) \}
    \end{displaymath}
    is called the {\bf admissible domain of $\gs$}.
    \end{defin}
    \begin{rem}\label{7D}
    The admissible domain $\kD(\gs)$ of a spectral family $\gs : \RR
    \to \kT(M)$ is dense in $M$.
    \end{rem}
    This follows directly from the observation that $U \cap
    \kD(\gs) = \emptyset$ for some $U \in \kT(M)$ implies $U \tm
    \bigwedge_{\gl \in \RR}\gs(\gl) = \emptyset$.

    On the other hand it may happen that $\kD(\gs) \ne M$. The
    spectral family $\gs_{ln}$ is a simple example:
    \begin{displaymath}
    	\forall \ \gl \in \RR : \ 0 \in \gs_{ln}(\gl).
    \end{displaymath}

    Each spectral family $\gs : \RR \to \kT(M)$ induces a function
    $f_{\gs} : \kD(\gs) \to \RR$:
    \begin{defin}\label{7E}
    Let $\gs : \RR \to \kT(M)$ be a spectral family with admissible
    domain $\kD(\gs)$. Then the function $f_{\gs} : \kD(\gs) \to
    \RR$, defined by
    \begin{displaymath}
    	\forall \ x \in \kD(\gs) : \ f_{\gs}(x) := \inf \{\gl \in \RR
    	\mid x \in \gs(\gl) \},
    \end{displaymath}
    is called the {\bf function induced by $\gs$}.
    \end{defin}
    In complete analogy to the operator case we define the spectrum of
    a spectral family $\gs$:
    \begin{defin}\label{7F}
    Let $\gs : \RR \to \kT(M)$ be a spectral family. Then
    \begin{displaymath}
    	R(\gs) := \{\gl \in \RR \mid \gs \  \text{is constant on a
    	neighborhood of}\  \gl \}
    \end{displaymath}
    is called the {\bf resolvent set} of $\gs$, and
    \begin{displaymath}
    	Spec(\gs) := \RR \setminus R(\gs)
    \end{displaymath}
    is called the {\bf spectrum} of $\gs$.
    \end{defin}
    Obviously $Spec(\gs)$ is a closed subset of $\RR$.
    \begin{prop}\label{7G}
    Let $f_{\gs} : \kD(\gs) \to \RR$ be the function induced by the
    spectral family $\gs : \RR \to \kT(M)$. Then
    \begin{displaymath}
    	Spec(\gs) = \overline{im f_{\gs}}.
    \end{displaymath}
    \end{prop}
    The functions induced by our foregoing examples are
    \begin{eqnarray}
    	f_{\gs_{id}}(x) & = & x
    	\label{1}  \\
    	f_{\gs_{abs}}(x) & = & |x|
    	\label{2}  \\
    	f_{\gs_{ln}}(x) & = & \ln{|x|} \quad  \text{and} \ \kD(\gs_{ln}) = \RR
    	\setminus\{0\}
    	\label{3}  \\
    	f_{\gs_{step}} & = & \sum_{n \in \ZZ}n\chi_{[n, n + 1[}
    	\label{4}
    \end{eqnarray}
    There is a fundamental difference between the spectral families
    $\gs_{id}, \gs_{abs}, \gs_{ln}$ on the one side and $\gs_{step}$
    on the other. The function induced by $\gs_{step}$ is not
    continuous. This fact is mirrored in the spectral families: the
    first three spectral families have the property
    \begin{displaymath}
    	\forall \ \gl < \mu : \ \overline{\gs(\gl)} \tm \gs(\mu).
    \end{displaymath}
    Obviously $\gs_{step}$ fails to have this property.
    \begin{defin}\label{7H}
    A spectral family $\gs : \RR \to \kT(M)$ is called {\bf
    continuous} if
    \begin{displaymath}
    	\forall \ \gl < \mu : \ \overline{\gs(\gl)} \tm \gs(\mu)
    \end{displaymath}
    holds.
    \end{defin}
    Using the pseudocomplement $U^{c} := M \setminus \bar{U} \quad (U
    \in \kT(M))$ we can express the condition of continuity in purely
    lattice theoretic terms as
    \begin{displaymath}
    	\forall \ \gl < \mu : \ \gs(\gl)^{c} \cup \gs(\mu) = M.
    \end{displaymath}
    \begin{rem}\label{7I}
    The admissible domain $\kD(\gs)$ of a continuous spectral
    family $\gs : \RR \to \kT(M)$ is an open (and dense) subset of $M$.
    \end{rem}
    \begin{rem}\label{7J}
    If $\gs : \RR \to \kT(M)$ is a continuous spectral family, then for
    all $\gl \in \RR \qquad \gs(\gl)$ is a regular open set, i.e.
    \begin{displaymath}
    	\gs(\gl)^{cc} = \gs(\gl).
    \end{displaymath}
    \end{rem}
    The importance of continuous spectral families becomes manifest
    in the following
    \begin{theo}\label{7K}
    Let $M$ be a toplogical space. Then every continuous function $f :
    M \to \RR$ induces a continuous spectral family $\gs_{f} : \RR
    \to \kT(M)$ by
    \begin{displaymath}
    	\forall \ \gl \in \RR : \ \gs_{f}(\gl) := int
    	\overset{-1}{f}(]-\infty, \gl]).
    \end{displaymath}
    The admissible domain $\kD(\gs_{f})$ equals $M$ and the function
    $f_{\gs_{f}} : M \to \RR$ induced by $\gs_{f}$ is $f$.
    Conversely, if $\gs : \RR \to \kT(M)$ is a continuous spectral
    family, then the function
    \begin{displaymath}
    	f_{\gs} : \kD(\gs) \to \RR
    \end{displaymath}
    induced by $\gs$ is continuous and the induced spectral family
    $\gs_{f_{\gs}}$ in $\kT(\kD(\gs))$ is the restriction of $\gs$ to
    the admissible domain $\kD(\gs)$:
    \begin{displaymath}
    	\forall \gl \in \RR : \ \gs_{f_{\gs}}(\gl) = \gs(\gl) \cap
    	\kD(\gs).
    \end{displaymath}
    \end{theo}
    The proof is, although not trivial, an exercise in general
    topology.

    Note that {\bf any} function $f : M \to \RR$ induces a spectral
    family $\gs_{f} : \RR \to \kT(M)$ by
    \begin{displaymath}
    	\gs_{f}(\gl) := int \overset{-1}{f}(]-\infty, \gl]).
    \end{displaymath}

    There is a close connection between bounded spectral families
    $\RR \to \LL(\kH)$ with values in a given Boolean sector $\BB \tm
    \LL(\kH)$ and spectral families \mbox{$\RR \to \kT\kQ(\BB))$.}\\
    Let $\gs : \RR \to \LL(\kH)$ be a bounded spectral family such
    that $\gs(\gl) \in \BB$ for all $\gl \in \RR$.\\
    For $\gl \in \RR$ define
    \begin{displaymath}
    	\hat{\gs}(\gl) := \bigwedge_{\mu > \gl}\kQ_{\gs(\mu)}(\BB) \in
    	\kT(\kQ(\BB)).
    \end{displaymath}
    \begin{prop}\label{7L}
    $\hat{\gs} : \RR \to \kT(\kQ(\BB))$ is a spectral family with
    admissible domain $\kQ(\BB)$.
    \end{prop}
    The connection between bounded spectral families $\RR \to
    \LL(\kH)$ with values in $\BB$ and spectral families $\RR \to
    \kT(\kQ(\BB))$ rests on the following
    \begin{prop}\label{7M}
    Let $\gs : \RR \to \LL(\kH)$ be a bounded spectral family with
    values in the Boolean sector $\BB \tm \LL(\kH)$, $f_{\gs} :
    \kQ(\BB) \to \RR$ the function induced by $\gs$, $\hat{\gs} :
    \RR \to \kT(\kQ(\BB))$ the spectral family induced by $\gs$ and
    $f_{\hat{\gs}} : \kQ(\BB) \to \RR$ the function induced by
    $\hat{\gs}$. Then
    \begin{displaymath}
    	f_{\hat{\gs}} = f_{\gs}.
    \end{displaymath}
    \end{prop}
    From this and from theorem \ref{7K} we obtain the following
    interesting result:
    \begin{cor}\label{7N}
    Let $\gs : \RR \to \LL(\kH)$ be a bounded spectral family with
    values in the Boolean sector $\BB$. Then $f_{\gs} : \kQ(\BB) \to
    \RR$ is continuous if and only if the spectral family
    $\hat{\gs}:\RR \to \kT(\kQ(\BB))$ is continuous.
    \end{cor}
    By the way this shows that in general the simpler ``Ansatz''
    \begin{displaymath}
    	\tilde{\gs}(\gl) := \kQ_{\gs(\gl)}(\BB)
    \end{displaymath}
    cannot give a spectral family $\tilde{\gs}$.

    Now let $\tau :\RR \to \kT(\kQ(\BB))$ be a continuous spectral
    family and let $f_{\tau}$ be the continuous function induced by
    $\tau$. There is exactly one $A \in C^{*}(\BB)_{sa}$ such that
    $f_{\gs_{A}} = f_{\tau}$ where $\gs_{A}$ denotes the spectral
    family of $A$. Hence
    \begin{cor}\label{7O}
    Let $\tau : \RR \to \kT(\kQ(\BB))$ be a continuous spectral
    family. Then there is a unique hermitian operator $A \in
    C^{*}(\BB)$ such that
    \begin{displaymath}
    	\hat{\gs_{A}} = \tau
    \end{displaymath}
    where $\gs_{A}$ is the spectral family of $A$.
    \end{cor}
    \vspace{15mm}

    In the following $\frA$ is an abstract $\gs$-algebra. Thus $\frA$
    is not necessarily a sub-$\gs$-algebra of the power set of some
    set $M$.
    \begin{defin}\label{7P}
    We define a spectral family in the $\gs$-algebra $\frA$ as a
    mapping
    \begin{displaymath}
    	\gs : \RR \to \frA
    \end{displaymath}
    with the following properties:
    \begin{enumerate}
    	\item  [(1)] $\gs(\gl) \leq \gs(\mu)$ for $\gl < \mu$.

    	\item  [(2)] $\gs(\gl) = \bigwedge_{n \in \NN}\gs(\gl_{n})$ for
    	every sequence $(\gl_{n})_{n \in \NN}$ such that $\gl_{n}
    	\searrow \gl$.

    	\item  [(3)] $\bigwedge_{n \in \NN}\gs(\gl_{n}) = 0$ for every
    	unbounded monotonously decreasing sequence $(\gl_{n})_{n \in \NN}$
    	in $\RR$.

    	\item  [(4)] $\bigvee_{n \in \NN}\gs(\gl_{n}) = 1$ for every
    	unbounded monotonously increasing sequence $(\gl_{n})_{n \in \NN}$
    	in $\RR$.
    \end{enumerate}
    \end{defin}
    If $\frA$ is a sub-$\gs$-algebra of the power set $pot(M)$ of
    some non-empty set $M$ (that means that $\bigwedge_{n}U_{n} =
    \bigcap_{n}U_{n}, \ \bigvee_{n}U_{n} = \bigcup_{n}U_{n}$ etc.), we
    define as in the topological case
    \begin{defin}\label{7Q}
    Let $\frA \tm pot(M)$ a sub-$\gs$-algebra and $\gs : \RR \to \frA$
    a spectral family. Then the function $f_{\gs} : M \to \RR$,
    defined by
    \begin{displaymath}
    	f_{\gs}(x) := \inf \{\gl \in \RR \mid x \in \gs(\gl) \},
    \end{displaymath}
    is called the function induced by $\gs$.
    \end{defin}
    Obviously we have
    \begin{rem}\label{7R}
    Let $\gs : \RR \to \frA$ be a spectral family in the
    sub-$\gs$-algebra \mbox{$\frA \tm pot(M)$.} Then
    \begin{displaymath}
    	\forall \ \gl \in \RR : \ \overset{-1}{f_{\gs}}(]-\infty,
    	\gl]) = \gs(\gl).
    \end{displaymath}
    \end{rem}
    \begin{cor}\label{7S}
    Let $\frA$ be as above and $\gs$ a spectral family in $\frA$.
    Then the function $f_{\gs} : M \to \RR$ is $\frA$-measurable. (We
    always assume that $\RR$ is equipped with the $\gs$-algebra of
    Borel sets.)
    \end{cor}
    Conversely, every $\frA$-measurable function $f : M \to \RR$
    defines a spectral family $\gs_{f} : \RR \to \frA$ by
    \begin{displaymath}
    	\gs_{f}(\gl) := \overset{-1}{f}(]-\infty, \gl]),
    \end{displaymath}
    and it can be easily seen that these constructions are inverse to
    each other:
    \begin{prop}\label{7T}
    Let $\frA \tm pot(M)$ be a sub-$\gs$-algebra. Then the spectral
    families $\RR \to \frA$ are in bijective correspondence to the
    $\frA$-measurable functions $M \to \RR$:
    \begin{displaymath}
    	\gs_{f_{\gs}} = \gs \quad \text{and} \quad  f_{\gs_{f}} = f.
    \end{displaymath}
    \end{prop}
    This result shows that the following definition of a {\bf real
    valued random variable for an arbitrary $\gs$-algebra $\frA$} is
    adequate:
    \begin{defin}\label{7U}
    Let $\frA$ be a $\gs$-algebra. An $\frA$-random variable is a
    spectral family $X :\RR \to \frA$.
    \end{defin}
    The consequences of this definition will be investigated in a
    forthcoming paper.


\begin{thebibliography}{99}

    \bibitem{birk} G. Birkhoff: Lattice theory,\\ AMS Coll. Publ.,
    1974
    \bibitem{ButIsh1}C.J. Isham and J. Butterfield: A topos
    perspective on the Kochen-Specker theorem: I. Quantum states as
    generalized valuations,\\ quant-ph/98 03 055
    \bibitem{ButIsh2} J. Butterfield and C.J. Isham: A topos
    perspective on the Kochen-Specker theorem: II. Conceptual
    aspects, and classical analogues,\\ quant-ph/98 08 067
    \bibitem{ButIshHam} J. Hamilton, C.J. Isham and J. Butterfield: A
    topos perspective on the Kochen-Specker theorem: III. Von Neumann
    algebras as the base category,\\ quant-ph/99 11 020
    \bibitem{CondeG} F. Constantinescu und H.F. de Groote:
    Geometrische und algebraische Methoden der Physik:
    Supermannigfaltigkeiten und Virasoro-Algebren,\\ Teubner Verlag,
    Stuttgart, 1994
    \bibitem{Jauch} J.M. Jauch: Foundations of quantum mechanics,\\
    Addison-Wesley, Reading, Mass., 1958
    \bibitem{MalRap} A. Mallios and I. Raptis: Finitary spacetime
    sheaves of quantum causal sets: Curving quantum causality,\\
    gr-qc/01 02 097
    \bibitem{PenRind} R. Penrose and W. Rindler: Spinors and
    space-time I, II,\\Cambridge Univ. Press, 1999
    \bibitem{Rap} I. Raptis: Finitary spacetime sheaves,\\ gr-qc/ 01
    02 108
    \bibitem{Sik} R. Sikorski: Boolean algebras,\\ Springer Verlag,
    1964
    \bibitem{stone} M.H. Stone: The theory of representations for
    Boolean algebras,\\ Trans. AMS 40 (1936), 37-111
    \bibitem{WardWells} R.S. Ward and R.O. Wells, Jr.: Twistor
    geometry and field theory,\\ Cambridge Univ. Press, 1995

    \end{thebibliography}
    \end{document}